\UseRawInputEncoding

\documentclass[draft]{agujournal2019}
\usepackage{url} 
\usepackage{lineno}
\usepackage[inline]{trackchanges} 
\usepackage{amsmath}
\usepackage{soul}
\graphicspath{{./figs_rev/}}
\makeatletter
\def\input@path{{./figs_rev/}}
\makeatother
\newcommand{\ig}[2]{\includegraphics[width = #1]{#2}}
\renewcommand{\vec}[1]{\mathbf{#1}}
\nolinenumbers
\pdfoutput=1
\draftfalse

\journalname{Journal of Geophysical Research: Space Physics}

\begin{document}

\title{Hybrid simulations of the cusp and dayside magnetosheath dynamics under quasi-radial interplanetary magnetic fields}

%
%




\authors{J. Ng \affil{1,2}, L.-J. Chen\affil{2}, Y. Omelchenko\affil{3,4}, Y. Zou\affil{5}, B. Lavraud\affil{6}}
\affiliation{1}{Department of Astronomy, University of Maryland, College Park, MD 20742, USA}
\affiliation{2}{NASA Goddard Space Flight Center, Greenbelt, MD 20771, USA}
\affiliation{3}{Trinum Research Inc., San Diego, CA 92126, USA}
\affiliation{4}{Space Science Institute, Boulder, CO 80301, USA}
\affiliation{5}{Department of Space Science, University of Alabama in Huntsville, Huntsville, AL, USA}
\affiliation{6}{Laboratoire d'astrophysique de Bordeaux, Univ. Bordeaux, CNRS, B18N, allée Geoffroy Saint-Hilaire, 33615 Pessac, France}
\correspondingauthor{Jonathan Ng}{jonng@umd.edu}







\begin{keypoints}
\item The effects of foreshock dynamics on the cusp and magnetosheath are studied by 3D hybrid simulations with different quasi-radial IMF. 
\item Under northward IMF, there is a depletion layer at the cusp-magnetosheath boundary, consistent with observations. 
\item Under southward IMF, foreshock turbulence and high-speed jets in concert with reconnection can lead to strong cusp density enhancements. 
\end{keypoints}

\begin{abstract}
Under quasi-radial interplanetary magnetic fields (IMF), foreshock turbulence can have an impact on the magnetosheath and cusps depending on  the location of the quasi-parallel shock. We perform three-dimensional simulations of Earth's dayside magnetosphere using the hybrid code HYPERS, and compare northward and southward quasi-radial IMF configurations. We study the magnetic field configuration, fluctuations in the magnetosheath and the plasma in the regions around the northern cusp. Under northward IMF with Earthward $B_x$, there is a time-varying plasma depletion layer immediately outside the northern cusp. In the southward IMF case, the impact of foreshock turbulence and high-speed jets, together with magnetopause reconnection, can lead to strong density enhancements in the cusp. 
\end{abstract}

\section*{Plain Language Summary}
When the magnetic field  in the solar wind is strongly Sun-Earth aligned, turbulence can develop in the region upstream of the Earth's magnetosphere and propagate towards the Earth. This turbulence has an impact on the magnetosheath and cusps. We use hybrid simulations with particle ions and fluid electrons to compare situations with a strong Sun-Earth aligned magnetic field combined with a weaker northward or southward component. When the northward component is present, a layer with low density is found outside the cusp, while in the case with the southward component, the impact of the turbulence, together with magnetic reconnection can lead to strong density enhancements in the cusp.

\section{Introduction}

Upstream of the quasi-parallel regions of the Earth's bow shock, ultra-low frequency (ULF) waves are excited by streaming instabilities due to the presence of reflected ions \cite{gary:1991}. The nonlinear evolution of these waves leads to the formation of structures such as Short Large Amplitude Magnetic Structures (SLAMS) \cite{schwartz:1992,chen:2021}, spontaneous hot flow anomalies \cite{zhang:2013}, cavitons \cite{kajdivc:2011} and high-speed jets \cite{hietala:2012,raptis:2020}. These structures can affect the cusp, magnetosheath and magnetopause, and contribute to turbulence  downstream of the quasi-parallel shock.

While the magnetosphere under radial IMF conditions has been studied using MHD simulations \cite{tang:2013}, it is important to capture the kinetic effects associated with the backstreaming ions in the foreshock.  Hybrid models with kinetic ions and fluid electrons (e.g.~ \cite{palmroth:2015,kempf:2015,omelchenko:2021}) are used to simulate these physical phenomena at reduced computational costs compared to fully kinetic simulations. With radial IMF configurations, these studies have shown the effects of ULF waves in the foreshock, associated structures such as cavitons  \cite{blancocano:2009,lin:2005}, turbulence downstream of the shock \cite{karimabadi:2014}, and high-speed jets in the magnetosheath \cite{ng:2021,omelchenko:2021jet,chen:2021jet}.

One area of interest is the magnetospheric cusp, the high-latitude funnel-shaped region through which solar wind plasma can directly access the ionosphere \cite{heikkila:1971,frank:1971}. The cusp has been the subject of numerous observational (e.g.~\citeA{lavraud:2005,pitout:2009,taylor:2004,nykyri:2012}) and simulation studies \cite{wang:2009,omidi:2007,palmroth:2013} under different IMF conditions. Under IMF with a strong radial component, the cusp has been observed to be highly dynamic \cite{taylor:2004}, and it has been shown that the turbulence in the ion foreshock can lead to ULF waves on the open and closed field lines of the polar cap \cite{shi:2020}. Other statistical observations have shown the structure of the cusp and its boundaries under various IMF conditions \cite{lavraud:2004,lavraud:2005,pitout:2012}, and statistical studies of cusp structure and motion are shown in \cite{pitout:2006,pitout:2009}. Energetic particles are found in the cusp, and studies have shown that they can originate from the magnetosphere \cite{delcourt:1999}, are accelerated locally \cite{nykyri:2012}, or are accelerated at the quasi-parallel shock \cite{trattner:2001}. Three-dimensional hybrid particle-in-cell simulations have been used, for instance, to trace the acceleration of cusp ions in the foreshock and the quasi-parallel shock \cite{wang:2009}. Another recent two-dimensional hybrid-Vlasov simulation work has studied proton precipitation in the cusps under purely northward and southward IMF, showing the importance of flux-transfer events in the southward case, and lobe reconnection in the northward case \cite{grandin:2020}.

In this work, we use three-dimensional hybrid simulations to study the dynamics of the cusp and dayside magnetosheath under quasi-radial northward and southward IMF conditions. We focus on comparing effects of foreshock turbulence on the overall structure of the dayside magnetosheath and boundaries of the northern exterior cusps in the two simulations.

\section{Simulation Model}
\label{sec:model}

In this study we use a hybrid particle-in-cell simulation code, HYPERS \cite{omelchenko:2012,omelchenko:2021,omelchenko:2021jet}. Ions are evolved kinetically as macroparticles while the electrons are treated as a charge-neutralising fluid. The magnetic field is updated using Faraday's law and the electric field is governed by a generalised Ohm's law:

\begin{align}
  \vec{E} &= \frac{\vec{J}_e\times (\vec{B}_{\text{self}}+\vec{B}_{ext})}{en_e c} - \frac{\nabla p_e}{n_e e} + \eta \vec{J}, \\
  \nabla \times \vec{B}_{\text{self}} &= \frac{4\pi}{c}\vec{J}, \vec{J} = \vec{J}_i + \vec{J}_e, \\
  \frac{\partial \vec{B}_{\text{self}}}{\partial t} &= -c\nabla\times \vec{E}, \\
  en_e &= \rho_i \\
  p_e &= n_e T_e \sim n_e^{\gamma}
\end{align}
Here $e$ is absolute value of the electron charge, $c$ the speed of light, $n_e$ the electron density, $\rho_i$ the ion charge density, $\vec{J}_{e,i}$ the electron and ion current densities, respectively, and $p_e$ the electron pressure with adiabatic constant $\gamma = 5/3$. $\vec{E}$ and $\vec{B}_{\text{self}}$ are the self-generated electric and magnetic fields and $\vec{B}_{ext}$ is a static magnetic field which includes the dipole field and the initial IMF, treated separately to minimize numerical errors. The magnetic field due to plasma currents is calculated self-consistently. 

In order to prevent the dipole field from affecting  the inflow boundary conditions, the simulations use a modified dipole formula which reduces the strength of the dipole field rapidly at large distances from the Earth \cite{omelchenko:2021jet}. The modified vector potential is

\begin{equation}
\vec{A}_{dip}(\vec{r}) = f(r) \frac{\vec{m}\times \vec{\hat{r}}}{r^2},
\end{equation}
where $\vec{m}$ is the magnetic dipole moment, $r = |\vec{r}|$, $\vec{\hat{r}} = \vec{r}/r$. We use $f(r) = \exp\left[-\left(r/r_{max}\right)^4\right]$ (note that $f(r) = 1$ corresponds to the standard dipole formula). We use $r_{max} = 450 d_i$, where $d_i$ is the ion inertial length using solar wind parameters, so that the modified field is very close to the standard dipolar field in all domains of interest. 

The resistivity $\eta$ used in this model is as follows:
\begin{equation}
\begin{split}
    \eta = \eta_{ch} + \eta_{v} \\
    \eta_{ch} = \frac{4\pi\nu_{ch}}{\omega_{pe}^2}, \nu_{ch} = c_{ch}\omega_{pi}\left[1-\exp\left(-\frac{v_d}{3v_s}\right)\right]\\
    \eta_{v} = \frac{4\pi\nu_{v}}{\omega_{pi,sw}^2}, \nu_{v} = c_{v}\omega_{ci,sw}\exp\left(-\frac{2n_e}{n_{min}}\right)\\
    v_d = |\vec{J}|/en_e, v_s = \sqrt{\gamma T_e/m_i}.
\end{split}
\end{equation}

Here $\omega_{pe}$ and $\omega_{pi}$ are the electron and ion plasma frequencies respectively, $n_{min}$ is the cutoff (minimum) electron density allowed in the simulation, $\vec{J}$ is the current density, $T_e$ is the electron temperature and $m_i$ is the ion mass. $\omega_{ci,sw}$ and $\omega_{pi,sw}$ are the initial solar-wind values of the respective quantities. The quantities $\nu_{ch}$ and $\nu_{v}$ are effective collision frequencies for the plasma and vacuum respectively in this model. 
$\eta_{ch}$ is the Chodura resistivity, which is an empirical expression  previously used to model field-reversed configurations \cite{omelchenko:2015}. The ``vacuum resistivity'' $\eta_v$ vanishes at cells occupied with plasma. 
The simulation parameters are chosen to be small enough not to significantly alter wave dispersion in the regions of interest: $c_{ch} = 0.01$, $c_v = 0.5$, $n_{min}=0.1 n_0$, where $n_0$ is the solar wind plasma electron number density.

Two hybrid simulations are performed with two different IMF directions and all other parameters remaining the same. The IMF is quasi-radial in both simulations with cone angles  $\pm 10^\circ$, with the $B_x$ component of the IMF pointing towards the Earth. The simulations use a physical domain of 716$d_i\times$1334$d_i\times$1334$d_i$, where $d_i$ is the ion inertial length in the solar wind. The upstream ratio between thermal and magnetic pressure for both ions and electrons is $\beta =  8\pi n_0 T_0/B_0^2 = 0.5$, and the ratio of the speed of light to the upstream Alfv\'en speed is $c/v_A = 8000$. Here $n_0$, $B_0$ and $T_0$ are the solar wind values of density, magnetic field and temperature respectively. The Earth's magnetic dipole is tilted 11.5$^\circ$ sunward. The IMF clock angle is zero. The computational domain is covered by a 300$\times$500$\times$500 cell stretched mesh, with a uniform patch around the Earth being covered by 175$\times$300$\times$300 grid cells with a resolution of 1 cell per $d_i$. Outside the uniform patch, the mesh cells are stretched exponentially with a growth factor of 12. For reference, the location of the uniform central patch is marked by the dashed lines in Fig.~\ref{fig:overall}. Simulations with similar parameters  have been discussed in \cite{ng:2021,omelchenko:2021jet,chen:2021jet}. The solar wind plasma is initialised with 25 particles per cell, which are split into two particles in the central domain. During the formation of the magnetosphere, the density increases by more than a factor of 4, so that the physical features of interest are resolved with $>200$ particles per cell.

We use the Geocentric Solar Magnetospheric (GSM) coordinate system, where the $x$-axis points from the Earth towards the Sun, the $y$-axis points in the dawn-dusk direction and the $z$-axis completes the right-handed system. Solar wind plasma is injected at the positive $x$ boundary of the simulation with velocity $v_x = -v_0$ and travels towards the magnetic dipole, which is centered at the $x=0$ boundary. The initial Alfv\'en Mach number is 8 (i.~e.~$v_0/v_A = 8$). The dipole strength is scaled such that the nominal magnetopause standoff distance is approximately $100$ $d_i$ (using the solar wind value of $d_i$). The inner boundary is a hemisphere of radius $50$ $d_i$ with absorbing boundary conditions for particles and perfectly conducting boundaries for fields. We note that distances are scaled down for computational feasibility such that $R_E \sim 12 d_i$ rather than realistically $\sim 60 d_i$ where $R_E$ is the Earth radius. We note that the standoff distance of $100$ $d_i$ is smaller than the realistic standoff distance ($\sim 500$ $d_i$), but is much larger than the minimum distance needed to simulate an earthlike magnetosphere \cite{omidi:2004}. Due to the reduced spatial scales, the turbulence scales may be exaggerated in this study, but the kinetic physics of shock formation, magnetic field dynamics and high-speed jets are consistent with observations \cite{omelchenko:2021jet}.

Unless otherwise mentioned, in the rest of the paper, $\rho = e n_e$ and $|\vec{B}|$ are normalised by their solar wind values, $\vec{v}$ is normalised by the solar wind Alfv\'en speed $v_{A,0}$, and lengths are normalised by the ion inertial length $d_i$. Temperature is normalised by $m_i v_{A,0}^2$ and $\Omega_{ci}$ is the ion cyclotron frequency calculated using the IMF.

\section{Results}

\begin{figure}
  \centering
  \ig{3.375in}{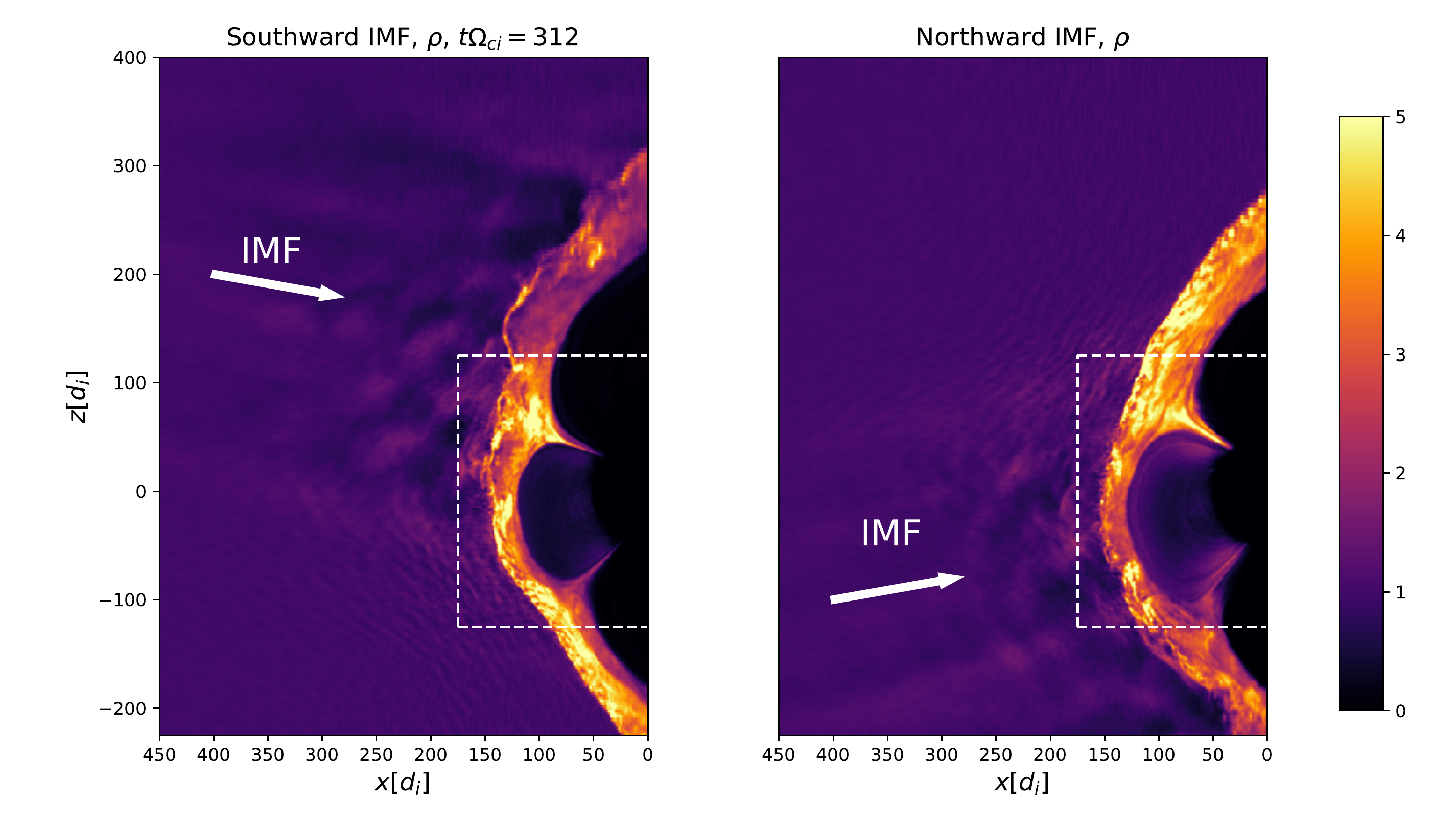}
  \caption{Overview of the two simulations at $t\Omega_{ci}$ = $312$ in the $x$-$z$ midplane. The dashed lines show the uniform central domain mentioned in Section~\ref{sec:model}. }
  \label{fig:overall}
\end{figure}

An overview of the two simulations is shown in Figure~\ref{fig:overall}, which shows the density $\rho$ in the central $x$-$z$ plane at the same simulation time after the magnetosphere has developed for both the southward and northward IMF cases. The panels show a subset of the entire domain, and the central domain with uniform grid spacing is marked by the dashed lines. In both cases, foreshock density fluctuations due to ULF waves can be seen upstream of the quasi-parallel regions of the bow shock consistent with prior work, with the location of the foreshock differing due to the IMF directions. Differences in the distributions of density, positions of the cusps and magnetopause are analysed in further detail below. There is also quasi-steady magnetopause reconnection in regions with $z < 0$ in the southward IMF case due to the favourable direction of the magnetic field. This has been discussed in another publication \cite{ng:2021}. The evolution of the system can be seen in movies in \cite{ng:2022movie}.

\subsection{Overall structure and fluctuations}
\label{sec:averaged}

In order to investigate the effects of turbulent structures in the foreshock, it is useful to look at time-averaged data in the simulation that illustrate its overall structure. For a physical quantity $Q$, we collect the data over 60 snapshots (with interval $3.125 \Omega_{ci}^{-1}$), and calculate the mean $\langle Q\rangle$ and standard deviation $\sqrt{\langle \delta Q^2\rangle}$ in each cell. In particular, we focus on the density $\rho$, magnetic field $|B|$ and ion temperature $T$.

\begin{figure}
  \centering
  \ig{3.375in}{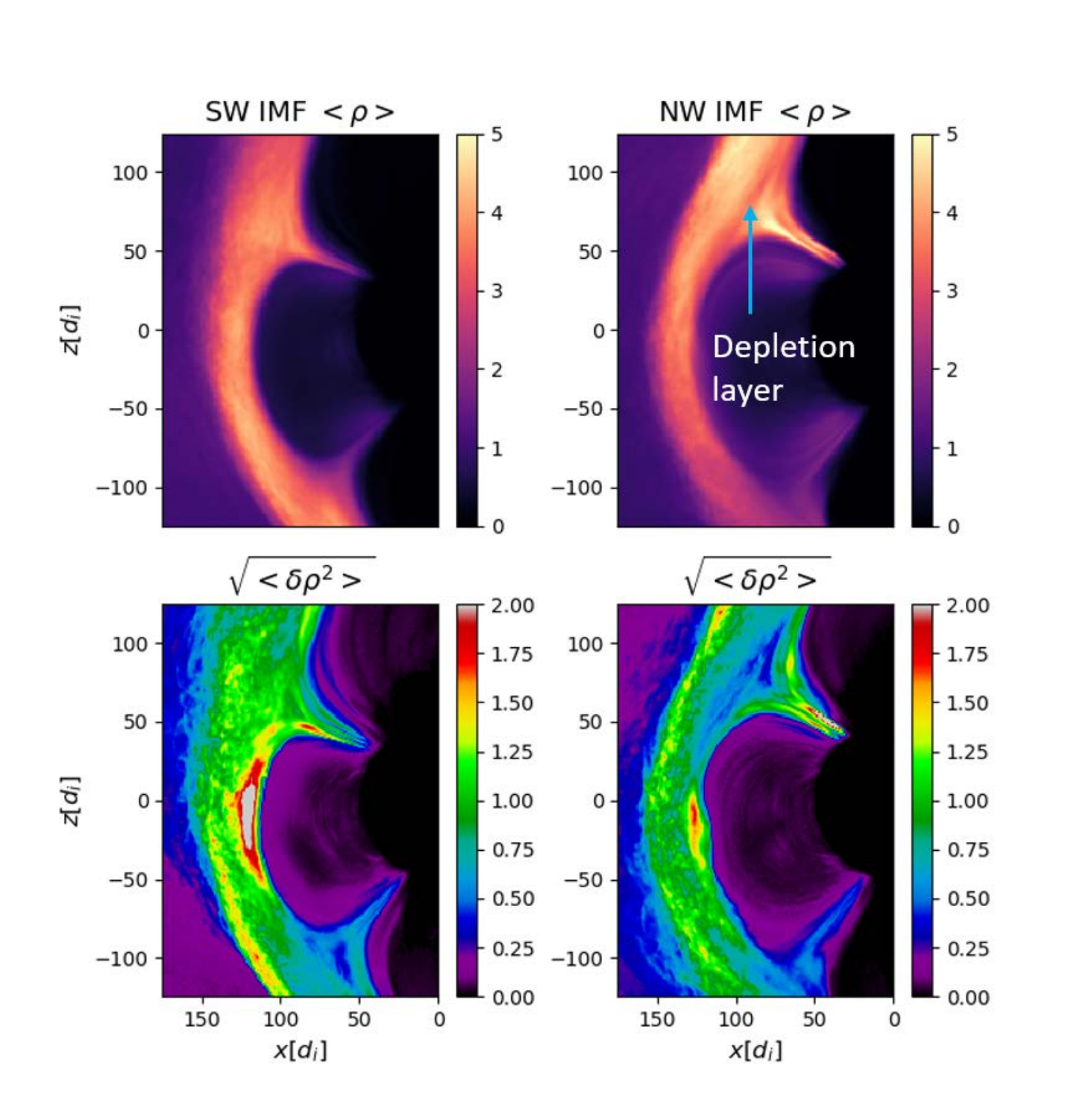}
  \caption{Time averaged density and standard deviation in the southward (left) and northward (right) IMF simulations. }
  \label{fig:rho}
\end{figure}

We first show the averaged density in the two simulations in Figure~\ref{fig:rho} in the $x$-$z$ plane at $y = 0.5$. In the top panels, immediate differences can be seen in the structure of the mean density. The average position of the bow shock and the magnetopause at the subsolar point is displaced outward by approximately $5 d_i$ ($\sim 0.4 R_E$) in the northward IMF case, and the density gradient across the magnetopause is gentler in the northward case. In the southward IMF simulation, the region of highest density in the magnetosheath is below $z = 0$, while the opposite is true for the northward case. The northward IMF case also shows a higher average density in the cusp region. Finally, in the regions close to the northern cusp, we note that in the southward IMF case, the structure of the average density remains continuous, while in the northward case, there appears to be a layer of reduced density forming a boundary between the magnetosheath and the cusp. This is  studied in greater detail  in Section~\ref{sec:ncusp}.  

With respect to the standard deviation of the density, the regions of most intense fluctuations are close to the magnetopause. While the solar wind is steady in the simulations, foreshock turbulence modifies its dynamic pressure (seen for a limited region in Fig.~\ref{fig:jet}, for example).

This can cause motion of the magnetopause which leads to the density fluctuations because the boundary between the high density magnetosheath and low density magnetospheric plasma is moving, causing spatial locations around the average magnetopause position to have varying density. Additionally, the magnetosheath is turbulent and there are frequent regions of strong density enhancements originating at the foreshock which necessarily terminate at the magnetopause. Some of these are associated with high-speed jets and can be seen in the movies \cite{ng:2022movie} and Fig.~\ref{fig:jet}. Large jets can cause indents as they impact the magnetopause \cite{omelchenko:2021jet,chen:2021jet}. There is also some contribution from flux-transfer events, though the quasi-radial IMF conditions are less favourable for their generation. 

Because of the strong radial IMF and dipole tilt, the quasi-parallel shock and foreshock extend to negative $z$ in the southward IMF case and positive $z$ in the northward IMF case, though the majority of the foreshock is in the positive and negative $z$ regions respectively (as seen in Fig.~\ref{fig:overall}). In the southward IMF case, strong transient density enhancements in the magnetosheath occur downstream of the quasi-parallel shock before being convected tailward in both the northward and southward directions. Together with the magnetopause motion, this leads to the broad region of strong density enhancement at the magnetopause seen in Fig.~\ref{fig:rho}. In the northward IMF case, the region of strong density fluctuations around $z \approx 0$ is likely to be caused by jets impinging on the magnetopause around $t\Omega_{ci} = 222, 272$ and the resulting density enhancement and magnetopause deformation. This can be seen in \cite{ng:2022movie}. 

In previous simulations, it has been shown that transient dayside magnetopause reconnection can be triggered even in the northward IMF case \cite{chen:2021jet}. The effects of the waves in the foreshock can also be seen in the regions upstream of the quasi-parallel regions of the bow shock, where the density fluctuations are larger. We also note in the northward IMF (Fig.~\ref{fig:rho}, lower-right panel) case that there is a quieter (blue) region in the northern cusp where the fluctuations are smaller, which may be related to the Stagnant Exterior Cusp \cite{lavraud:2002,lavraud:2004}.

\begin{figure}
  \centering
  \ig{5.375in}{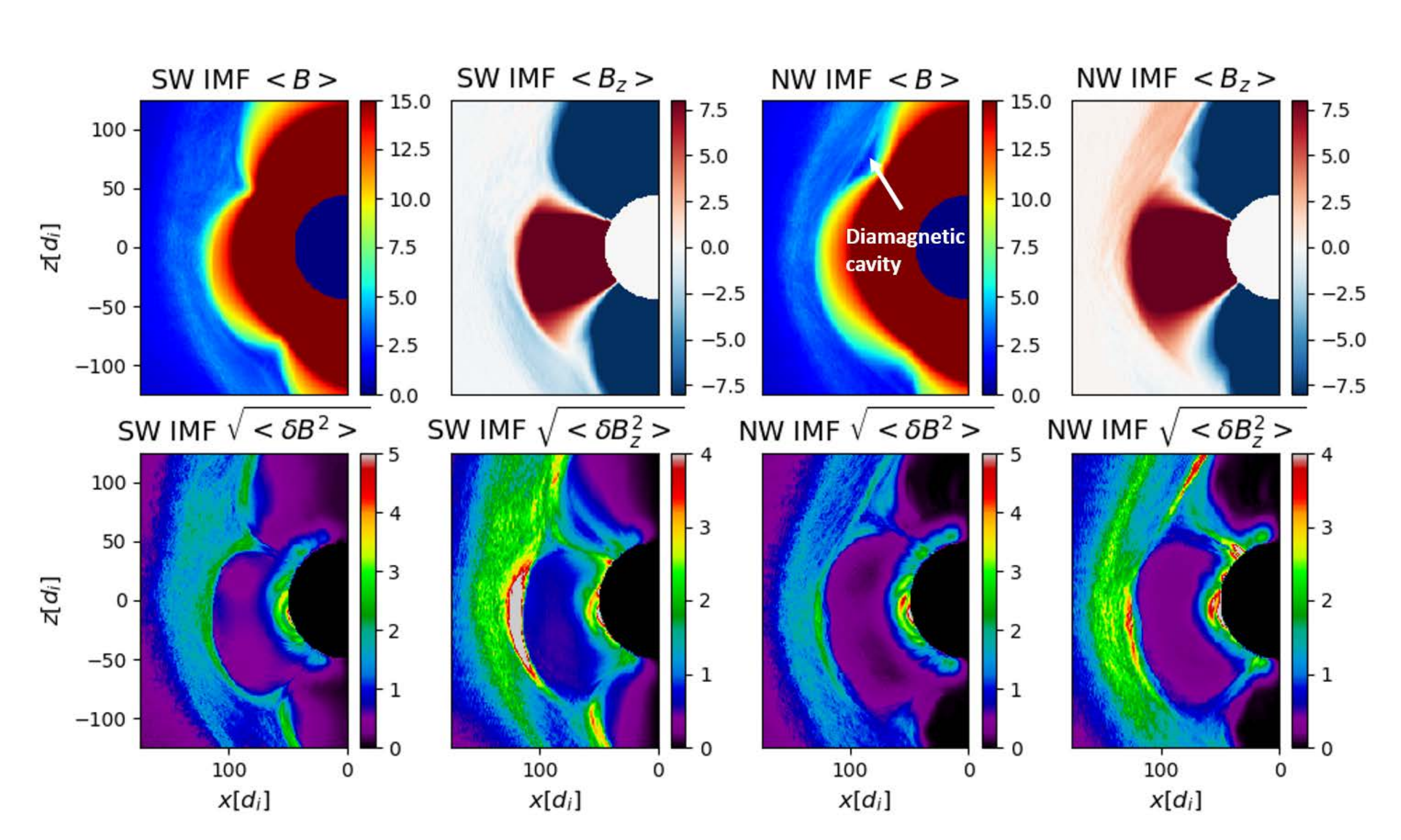}
  \caption{Time averaged magnetic field $|B|$ and $B_z$ and standard deviations in the southward (first and second columns) and northward (third and fourth columns) IMF simulations. }
  \label{fig:BT}
\end{figure}

In Figure~\ref{fig:BT} we show the magnitude of the magnetic field and its $z$-component. In both runs the magnetosheath fluctuations of $|B|$ are large ($ > 0.1 \langle B\rangle$), with the largest values found close to the magnetopause boundary. In the southward IMF case, there is also a region of weaker fluctuations at the southern cusp. The fluctuation level in the foreshock is much lower, consistent with the results of 2D hybrid simulations \cite{karimabadi:2014}. Another notable feature in the northward IMF simulation is a region of reduced $|B|$ at the northern cusp ($x\approx 90, z\approx 60$), indicative of the cusp diamagnetic cavity or exterior cusp \cite{lavraud:2002,nykyri:2011,burkholder:2021}. In both cases, although the averaged $B_z$ in the magnetosheath is relatively smooth, there are strong electromagnetic fluctuations that can be seen in the $\langle \delta B_z^2\rangle$ plots. These can also be seen in the movies in \cite{ng:2022movie}. There are also particularly strong $B_z$ fluctuations at the magnetopause in the southward IMF case in the same region as the density fluctuations shown in Fig.~\ref{fig:rho}. These are mainly caused by both the magnetopause motion and the transient intensification of the magnetic field just upstream of the magnetopause \cite{ng:2021}. The region of largest $\langle \delta B_z^2\rangle$  has a width of approximately 14 $d_i$ ($\sim 1.2 R_E$). In addition to the dipole tilt, which places the northern cusp further towards the dayside, the north-south asymmetry in fluctuations in both runs is also affected by the different locations of the quasi-parallel shock, consistent with observations of magnetic field perturbations under quasi-radial IMF conditions \cite{shi:2020}.

\begin{figure}
  \centering
  \ig{4.375in}{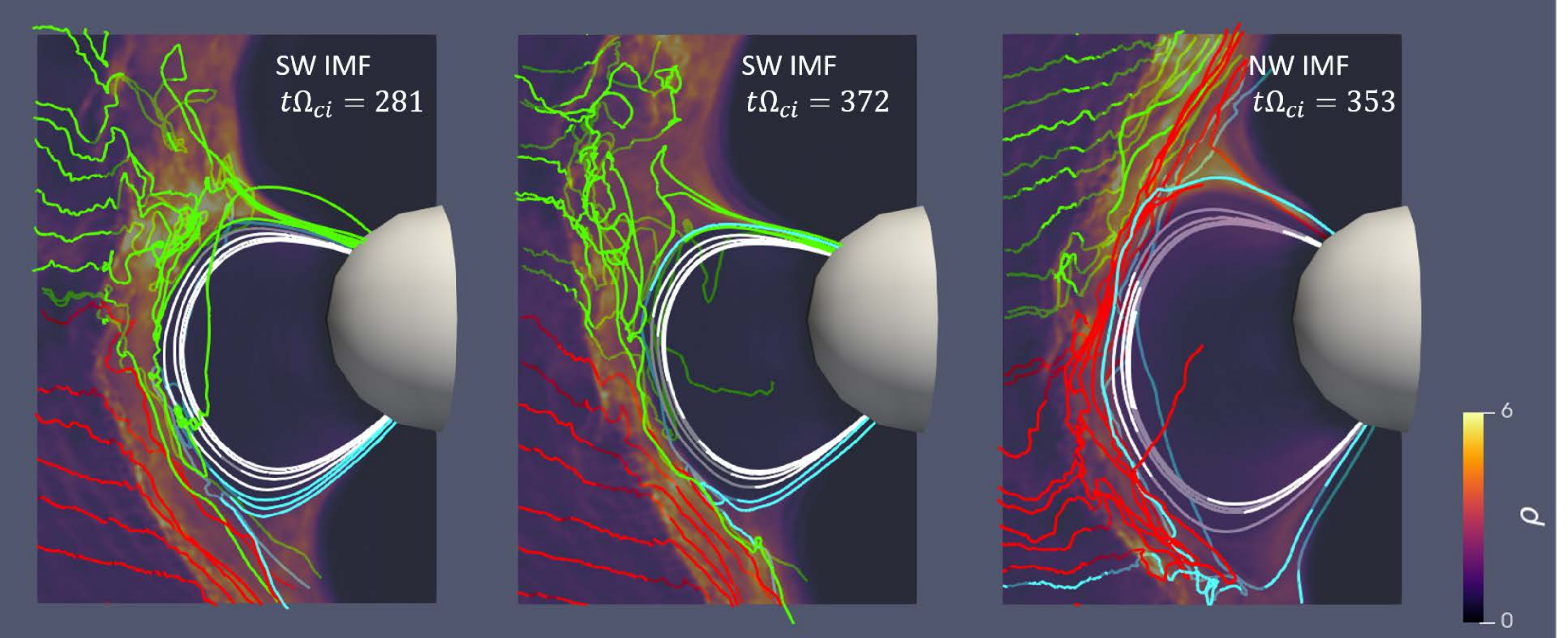}
  \caption{Magnetic field lines in the 3D simulations at selected times. The semi-transparent coloured $x$-$z$ plane is at $y=0$ and shows the density. The red and green lines are seeded in the IMF and magnetosheath at locations with $z <0$ and $z > 0$ respectively. White lines are seeded in the magnetosphere. Cyan lines are seeded close to the boundary between the magnetosheath and southern cusp. Field lines are three-dimensional, with the darkened regions behind the plane. }
  \label{fig:field_lines}
\end{figure}

The structure of magnetic field lines in both cases at selected times is shown in Fig.~\ref{fig:field_lines}. The left and middle panels show the magnetic field lines in the southward IMF simulation at two selected times which will be discussed later in this Section, while the right panel shows the northward IMF case. For visual clarity, different coloured field lines are used, with the red and green lines being seeded in the IMF and magnetosheath at locations with $z < 0$ and $z > 0$ respectively, while the closed  white field lines are seeded in the magnetosphere. Cyan field lines are seeded close to the boundary between the magnetosheath and southern cusp. Because of the quasi-radial IMF with a component towards the Earth, the magnetosheath field lines are generally pointing northward in the Northern Hemisphere, and southward in the Southern Hemisphere \cite{pi:2017}, with the exact location of where $B_z$ changes sign  depending on the simulation IMF conditions and local magnetosheath conditions. In the southward IMF case, there is a quasi-steady magnetopause reconnection region in the Southern Hemisphere due to the southward magnetic field \cite{ng:2021}, and in the Northern Hemisphere, some of the field lines in the turbulent magnetosheath are connected to the cusp \cite{omelchenko:2021jet}. In the northward IMF simulation, there is no steady reconnection at the subsolar magnetopause, though turbulence in the magnetosheath has been shown to cause dayside magnetopause reconnection in simulations \cite{chen:2021jet}. The field lines in the Northern Hemisphere can reconnect tailward of the cusp, and there is a clear boundary between the magnetosheath and cusp, which will be discussed in the next section.

\subsection{Northern cusp}
\label{sec:ncusp}

We now show results focusing on subsets of the global domain around the northern cusp at times which highlight key differences between the two simulations.

\begin{figure}
  \centering
  \ig{5.375in}{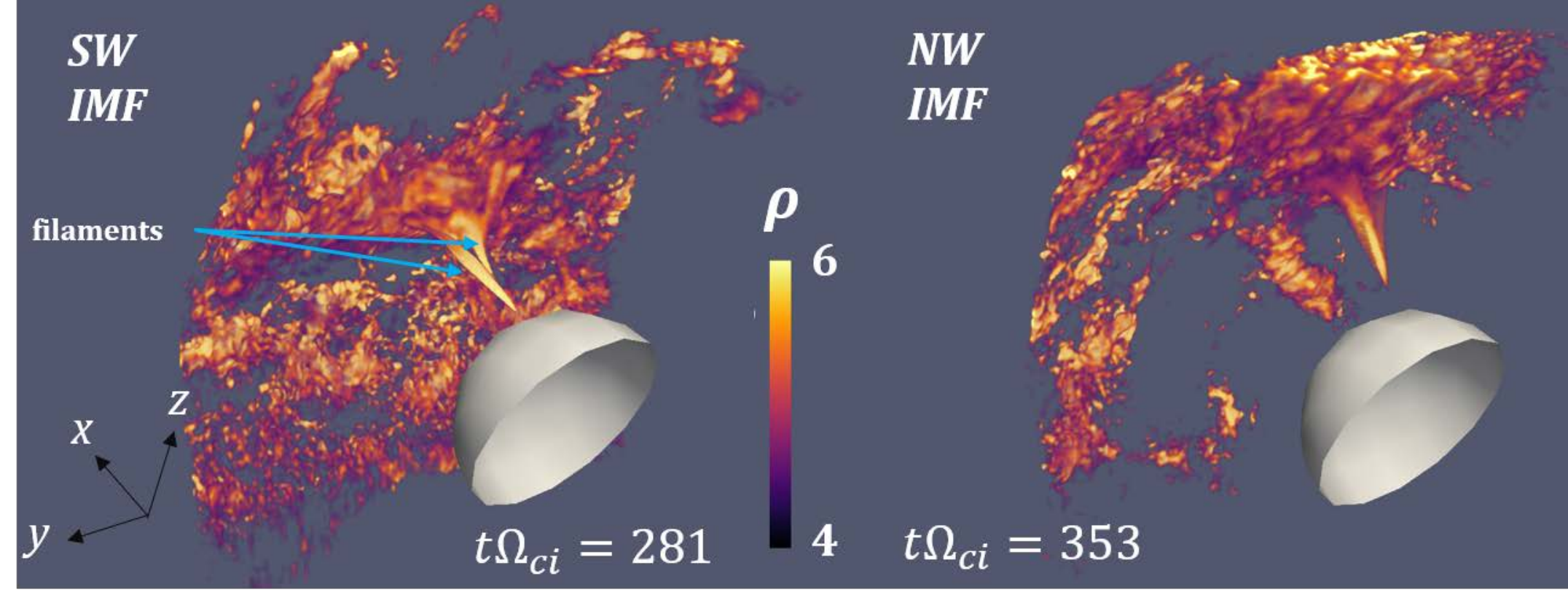}
  \caption{Volume rendering of the plasma density. Regions with $\rho < 4.2$ are transparent. }
  \label{fig:cusp3d}
\end{figure}

Selected snapshots from both simulations are shown in Figure~\ref{fig:cusp3d}. The left panel is taken from the southward IMF simulation, while the right panel is from the northward IMF simulation. In these figures, the volume rendering is such that regions with $\rho < 4.2$ are transparent, in order to emphasize  density enhancements. Filamentary density enhancements are present in the southward IMF case. This is evident in the left panel of Fig.~\ref{fig:cusp3d}. These filaments are connected to the magnetosheath by reconnected magnetic field lines, and the role of reconnection in their formation will be the subject of future study.

\begin{figure}
  \centering
  \ig{4.375in}{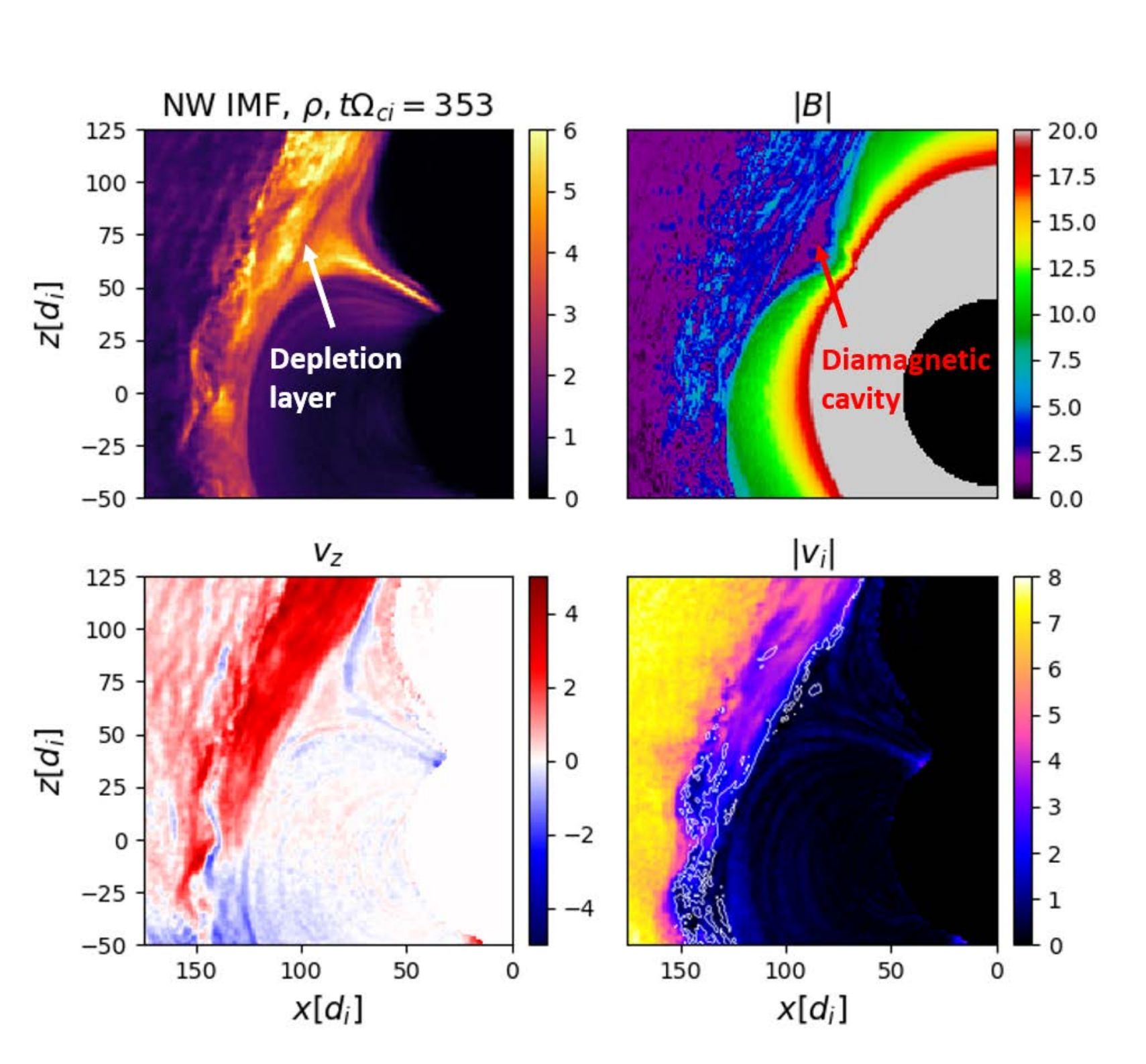}
  \caption{Density, magnetic field, $|B|$, velocity $v_z$ and total ion velocity at $t\Omega_{ci} = 353$ in the northward IMF simulation. Contours in the $|v_i|$ plot are at local $M_A = 1$. The range of values for $|B|$ is chosen to highlight the magnetosheath activity }
  \label{fig:panels_nw}
\end{figure}

\begin{figure}
  \centering
  \ig{4.375in}{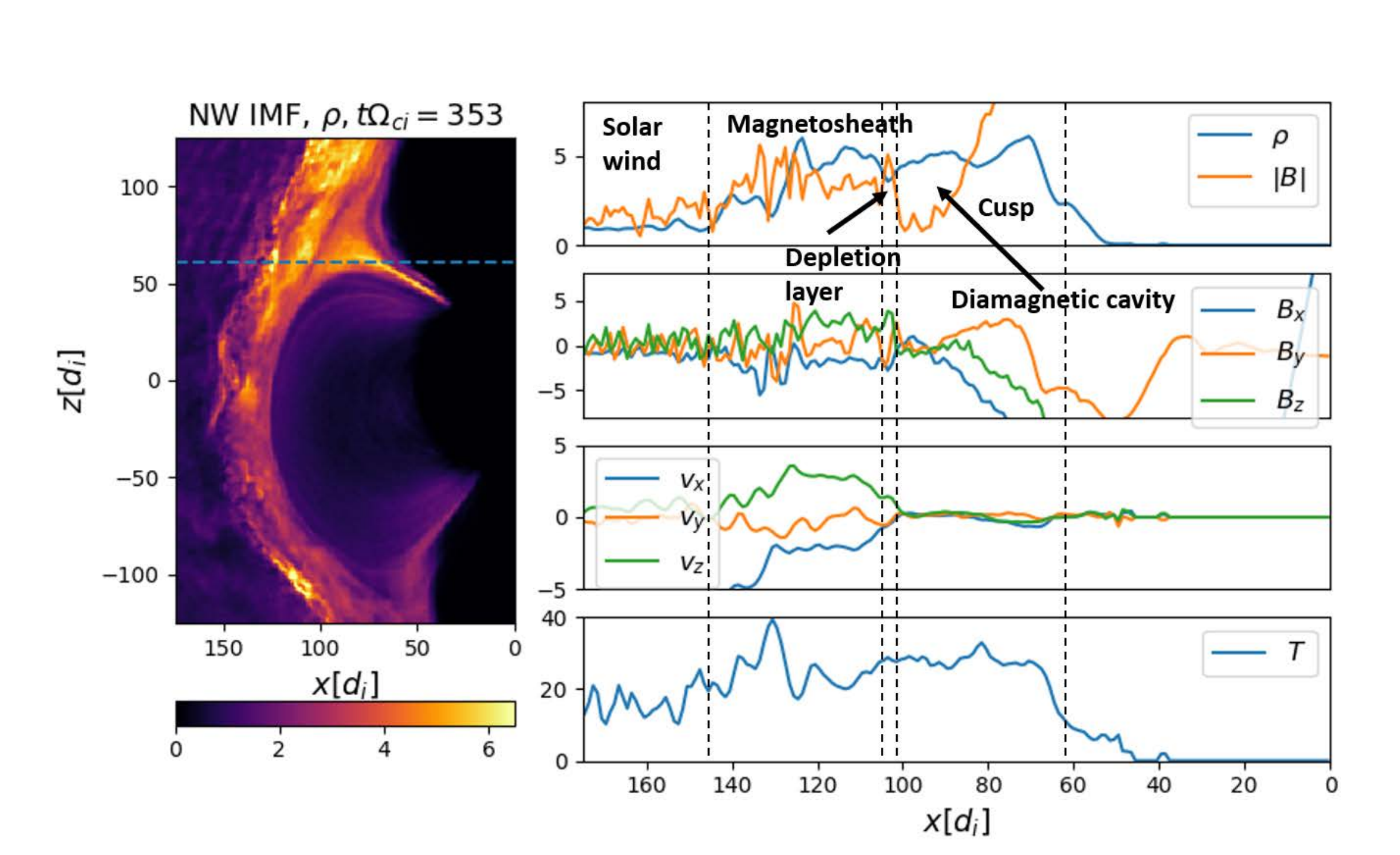}
  \caption{Left: Density in the northward IMF simulation. Dashed line shows cut along which other quantities are computed. Right: Density, magnetic field, velocity and scalar temperature along the cut. }
  \label{fig:cut_nw}
\end{figure}

Figure~\ref{fig:panels_nw} shows the density, magnetic field, $v_z$ and total velocity in the midplane of the northward IMF simulation. One of the more striking features is the presence of a clear boundary in the $\rho$ plot ($50 < z < 100$, $50 < x < 110$), where a region of reduced density can be seen between the cusp and the magnetosheath. This region shows an increased $|B|$ and a $v_z$ reversal. The total ion velocity also shows a clear boundary, which was used in \cite{lavraud:2004} to identify the cusp-magnetosheath boundary. The plasma flow is super-Alfv\'enic in the magnetosheath and reaches a sub-Alfv\'enic local minimum at the boundary layer, as shown by the contours of $M_A = 1$ in the $|v_i|$ plot. Here $M_A$ is calculated using local parameters. Past the boundary layer further into the cusp, there are transient and patchy super-Alfv\'enic regions. On average, flows within the cusp are sub-Alfv\'enic. The presence of these regions is likely caused by the low magnetic field and high density reducing the local Alfv\'en speed rather than strong flows. As can be seen in Figs.~\ref{fig:panels_nw} and \ref{fig:cut_nw}, the flow speed is much smaller in this region than in the magnetosheath. As mentioned earlier, there is a clear diamagnetic cavity ($75 < x < 100$), with $|B|$ increasing deeper in the cusp, as expected from observations \cite{lavraud:2002}. 

A one-dimensional cut through the boundary layer is shown in Figure~\ref{fig:cut_nw}. Here we focus on the boundary between the cusp and the magnetosheath plasma. In the topmost panel, we note a decrease in density by approximately 30\% from the magnetosheath to  $x = 103$ and a corresponding increase in $|B|$ by approximately 40\% compared to the magnetosheath region. This layer has a width of approximately $4 d_i$ and is located in the velocity shear layer where $v_x$ and $v_y$ vanish, and $v_z$ starts to decrease. There is a slight increase in ion temperature from the magnetosheath to the cusp. This region ($101<x<105$) is likely to be a plasma depletion layer (PDL) \cite{zwan:1976}, and observations have shown the presence of such layers at the cusp boundary under northward IMF (with various IMF $B_x$ and $B_z$) \cite{lavraud:2004}. The plasma depletion layer forms due to the northward magnetic fields piling up as they cannot reconnect. The increased magnetic pressure then causes  a reduction in plasma pressure and density, leading to the formation of a PDL \cite{zwan:1976,pi:2017}. The pile-up of the magneitc field lines can be seen in the right panel of Fig.~\ref{fig:field_lines}. In the simulation, the depletion layer is a dynamic structure, with a varying width and magnitude of the density drop. Over the preceding $30\Omega_{ci}^{-1}$, the width of the depletion layer varies between $4$ and $10$ $d_i$ in the $x$-direction, and the density decrease is approximately 20\% to 40\% of the magnetosheath density.  

\begin{figure}
  \centering
  \ig{4.375in}{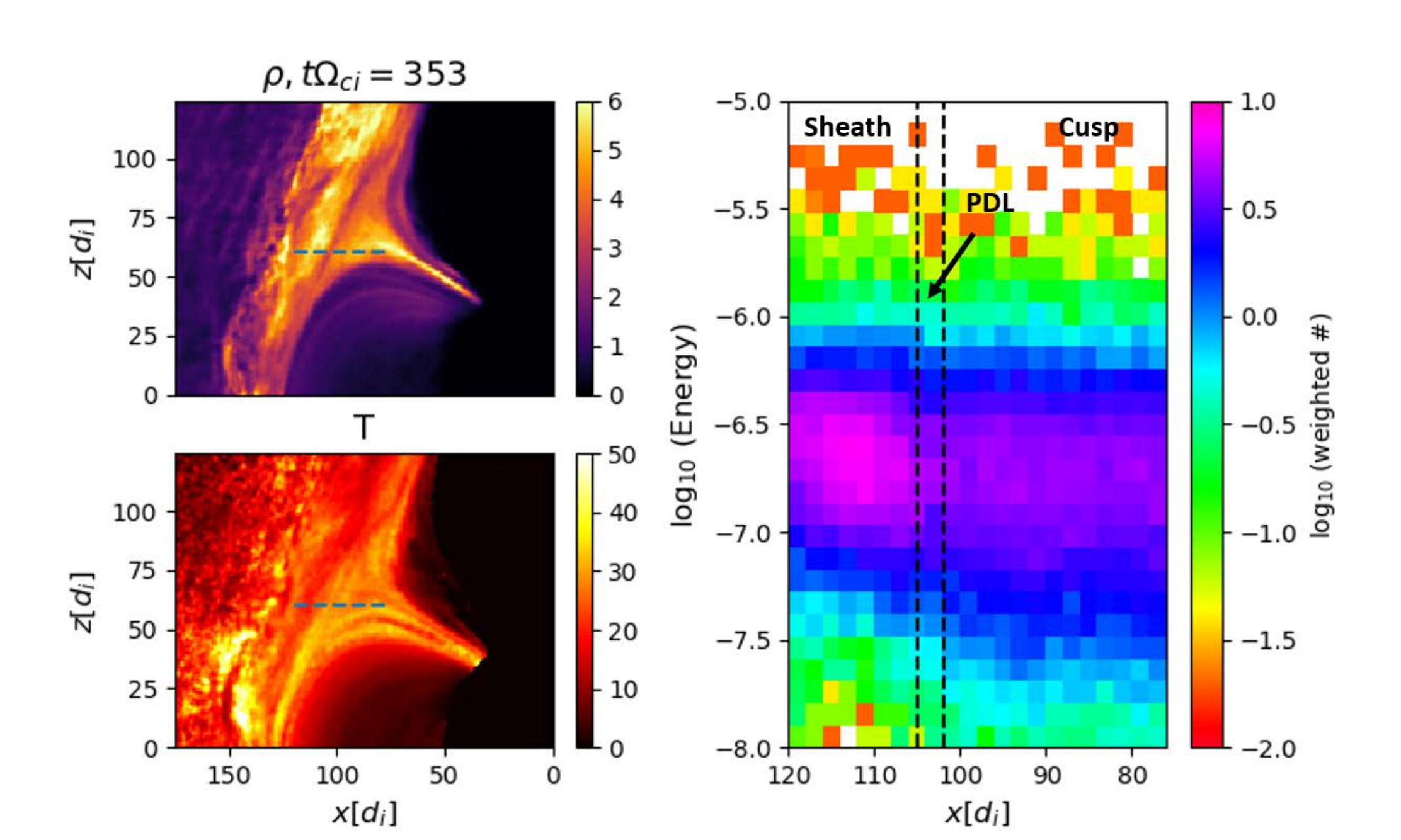}
  \caption{Left: Density and temperature in the northward IMF simulation. Dashed lines show the cut from which the particle distribution is taken. Right: Ion energy distribution. Units of energy are $m_i c^2$. }
  \label{fig:particle_nw}
\end{figure}

The ion energy distribution in this region can be seen in Figure~\ref{fig:particle_nw}.  In this Figure, the density and temperature are shown in the left panels as a guide, and particle data are taken from a region extending $\pm 5 d_i$ in the $y$ and $z$ directions along the dashed line. The transition between the magnetosheath and cusp can be seen in the energy spectrum around $x = 105$ (right panel), where a larger mean energy is seen in the magnetosheath, and a broader energy distribution with lower energy ions is found in the cusp.

\begin{figure}
  \centering
  \ig{4.375in}{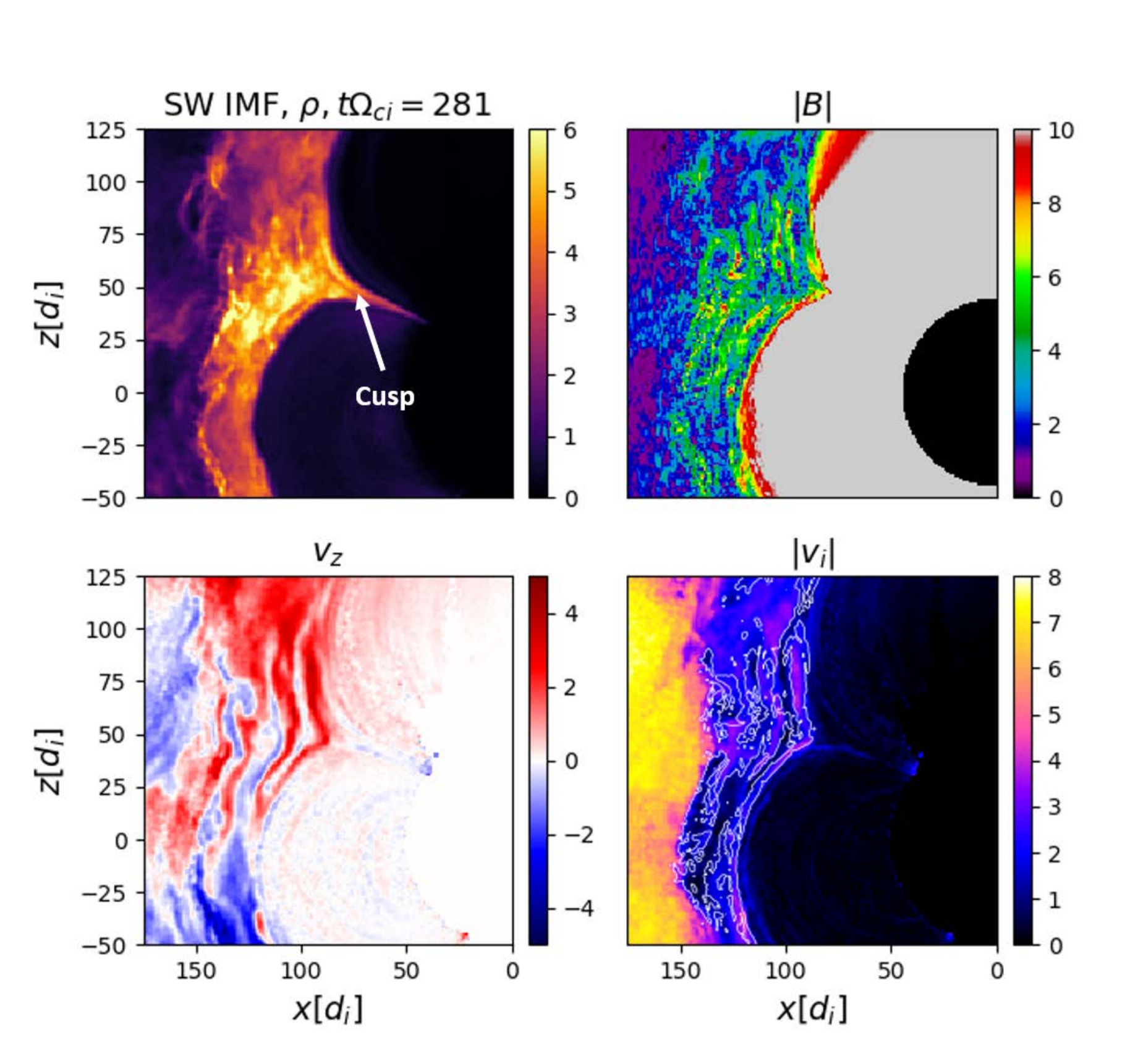}
  \caption{Density, magnetic field, $|B|$, velocity $v_z$ and total velocity  at $t\Omega_{ci} = 281$ in the southward IMF simulation. Contours in the $|v_i|$ plot show $M_A = 1$. }
  \label{fig:panels_sw}
\end{figure}

\begin{figure}
  \centering
  \ig{4.375in}{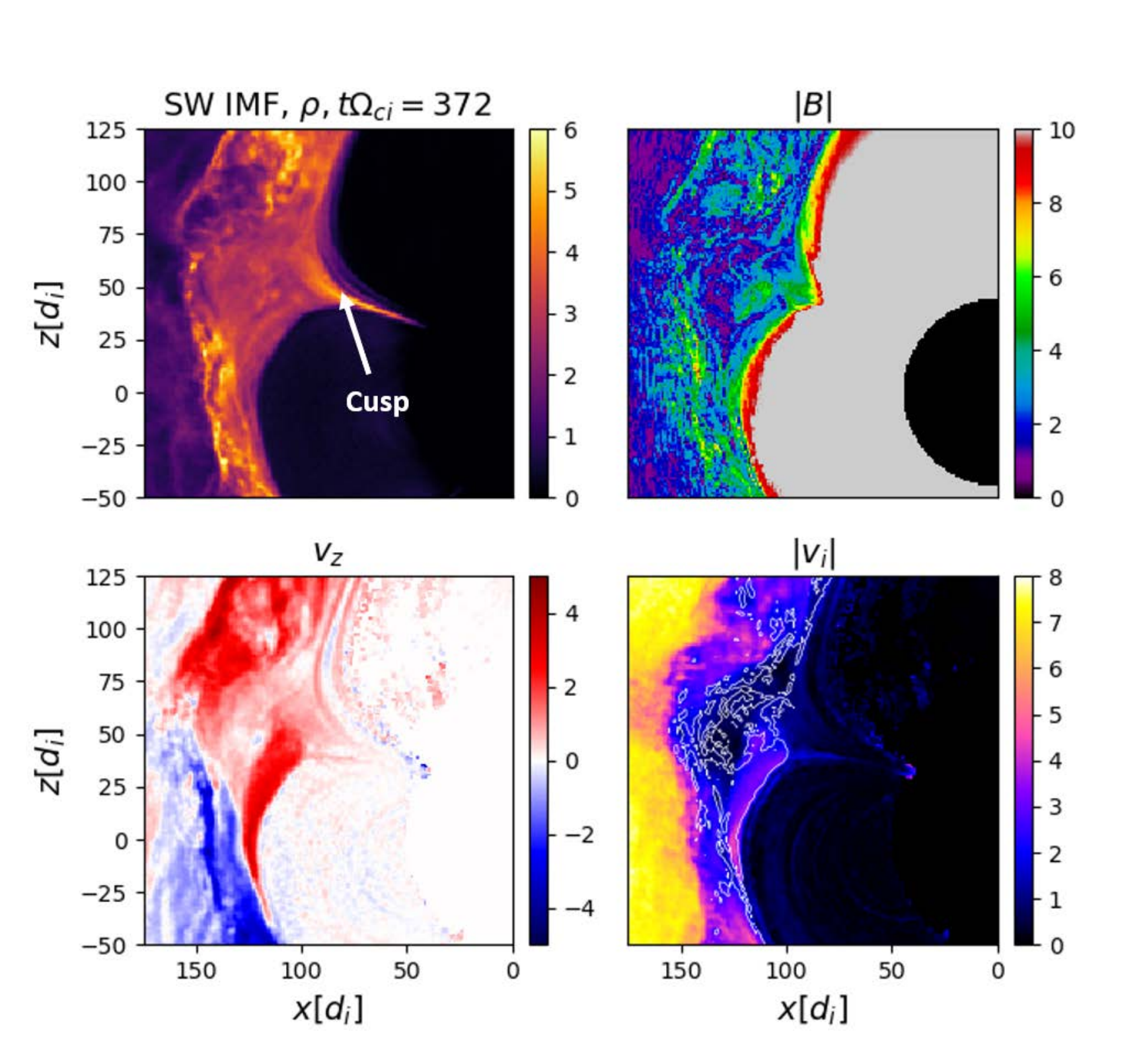}
  \caption{Density, magnetic field, $|B|$, velocity $v_z$ and total velocity at $t\Omega_{ci} = 372$ in the southward IMF simulation. Contours in the $|v_i|$ plot show $M_A = 1$.}
  \label{fig:panels_sw2}
\end{figure}

\begin{figure}
  \centering
  \ig{4.375in}{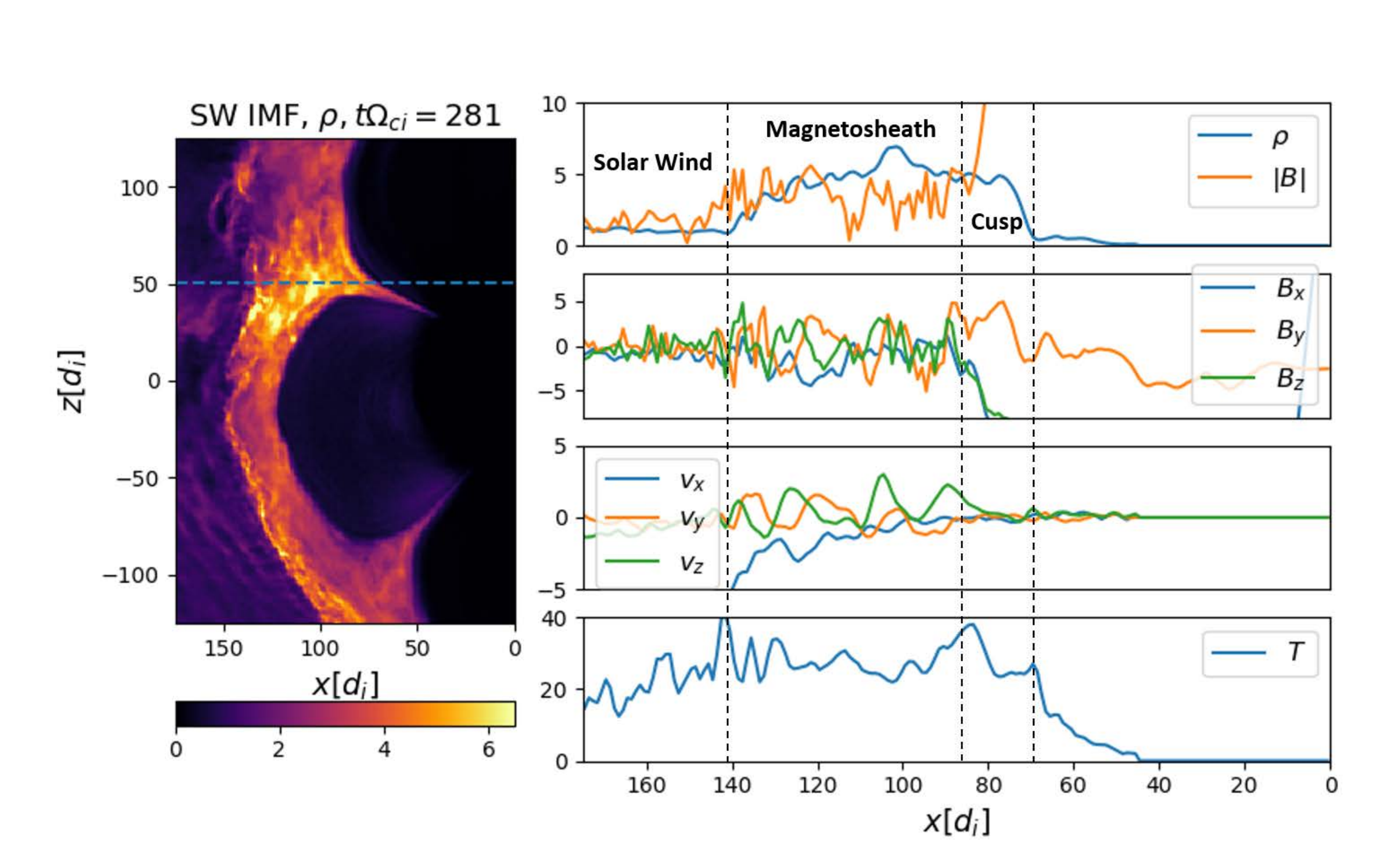}
  \caption{Left: Density in the southward IMF simulation at $t\Omega_{ci} = 281$ when there is a magnetosheath jet. Dashed line shows the cut along which other quantities are computed. Right: Density, magnetic field, velocity and scalar temperature along the cut. }
  \label{fig:cut_sw}
\end{figure}

We now study the case with southward IMF. At different times, the structures of the magnetosheath and the cusp are quite different due to the effects of foreshock turbulence on the magnetosheath. This can be seen in Figs.~\ref{fig:panels_sw} and~\ref{fig:panels_sw2}. At $t\Omega_{ci} = 281$, there is a strong density enhancement in the magnetosheath, together with variations in $|B|$ and $v_z$ at $z \approx 50$.  The variation of quantities along a one-dimensional cut can be seen in Fig.~\ref{fig:cut_sw}, in which the components of $B$ show strong fluctuations comparable to the mean magnetic field in the region. The evolution of the density over a longer time interval can also be seen in movies \cite{ng:2022movie}. 

\begin{figure}
  \centering
  \ig{4.375in}{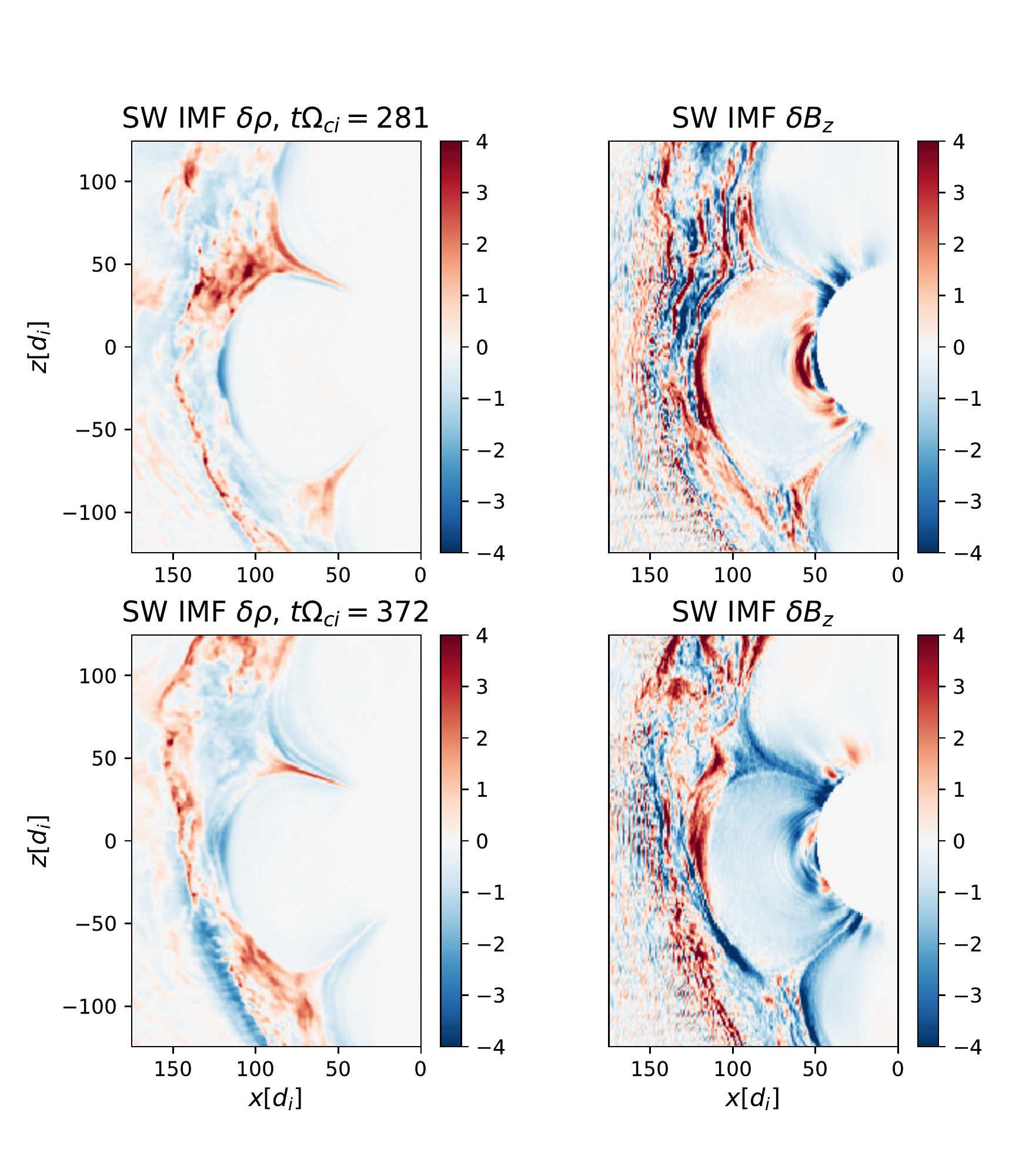}
  \caption{Deviations from the mean density and magnetic field $B_z$ at $t\Omega_{ci} = 281$ (top) and $t\Omega_{ci} = 372$ (bottom) in the southward IMF simulation.}
  \label{fig:fluctuations_sw}
\end{figure}

The strong density enhancement and magnetic field fluctuations at $t\Omega_{ci} = 281$ can be seen in Fig.~\ref{fig:fluctuations_sw}. Here the deviations of density $\rho$ and $B_z$ from their mean values are shown, and it can be seen that there are large density fluctuations ($\delta \rho$) in the region centred around $x \approx 100$, $z \approx 50$ comparable in size to the mean density in this region  (shown in Fig.~\ref{fig:rho}). There are also strong $B_z$ fluctuations correlated with the spatially oscillating $v_z$ pattern ($25 < z < 75$) seen in Fig.~\ref{fig:panels_sw}.

\begin{figure}
  \centering
  \ig{4.375in}{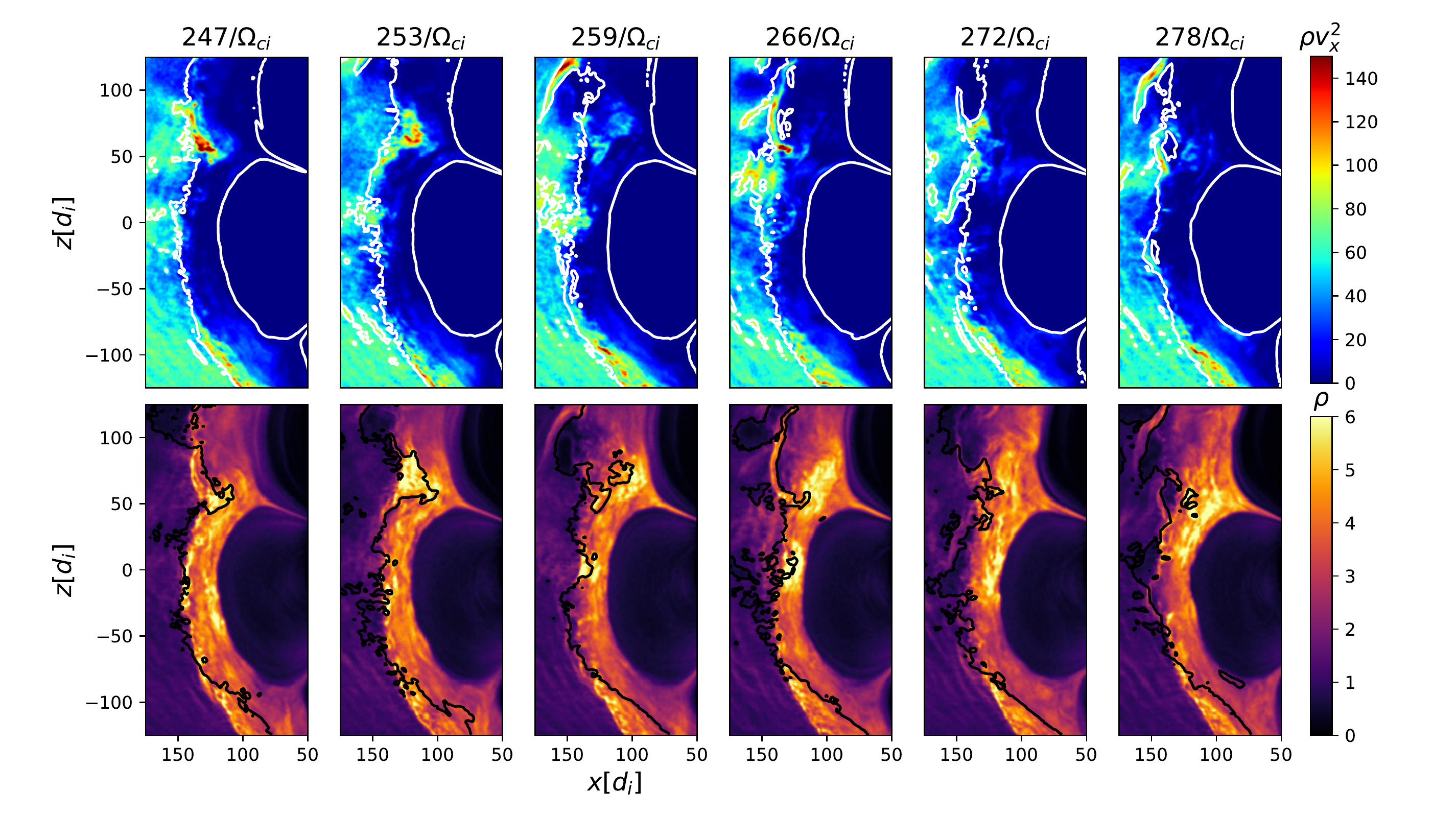}
  \caption{Evolution of high-speed jets in the southward IMF simulation. Top: Dynamic pressure $\rho v_x^2$ used to identify magnetosheath jets. White contours are of $\rho = 1.5$ and show the approximate locations of the bow shock, subsolar regions of the magnetopause and boundary that separates the plasma lobe from the cusp and magnetosheath. Bottom: Evolution of the density. Black contours are $0.5 \rho_{sw} v_{sw}^2$, half of the dynamic pressure in the solar wind. }
  \label{fig:jet}
\end{figure}

The effect of jets on the magnetosheath and cusp can be further seen  in Fig.~\ref{fig:jet}, where the time evolution of the density and dynamic pressure $\rho v_x^2$ over the preceding $34/\Omega_{ci}$ is shown. The top row shows the enhancement of the dynamic pressure which is used to identify magnetosheath jets \cite{plaschke:2018}, while the bottom row shows the density. Contours ($\rho = 1.5$) in the top row of panels show the approximate locations of the bow shock, subsolar magnetopause and the boundaries between the lobe, cusp and magnetosheath, while contours in the bottom row have the threshold value of half of the solar wind dynamic pressure and are used to identify jet positions. Here it can be seen that these jets cause strong magnetosheath density enhancements first seen at $t = 253/\Omega_{ci}$ at $z\approx 70$ and $t=259/\Omega_{ci}$ at $z\approx 0$. and there are still regions of high dynamic pressure at $z\approx 50$ in the last three panels. The jet dynamics initially leads to an increase in the density in the magnetosheath outside the cusp. As the system evolves, the region of increased density at $z\approx 0$ is convected upwards. Additional jets impact the region around $z=50$. The presence of the jet at $z=50$ means that there is a bulk plasma flow from the bow shock with dominant $v_x$ up to approximately $x=100$, leading to the compression of the plasma ahead of it. Reconnection is still important to this process as it allows the plasma from the region of enhanced density in the magnetosheath to access the cusp. As seen in Fig.~\ref{fig:field_lines}, many of the green the field lines in this region are connected to the IMF and northern cusp. 
The evolution of the density and dynamic pressure can also be seen in  movies at \cite{ng:2022movie}.

\begin{figure}
  \centering
  \ig{4.375in}{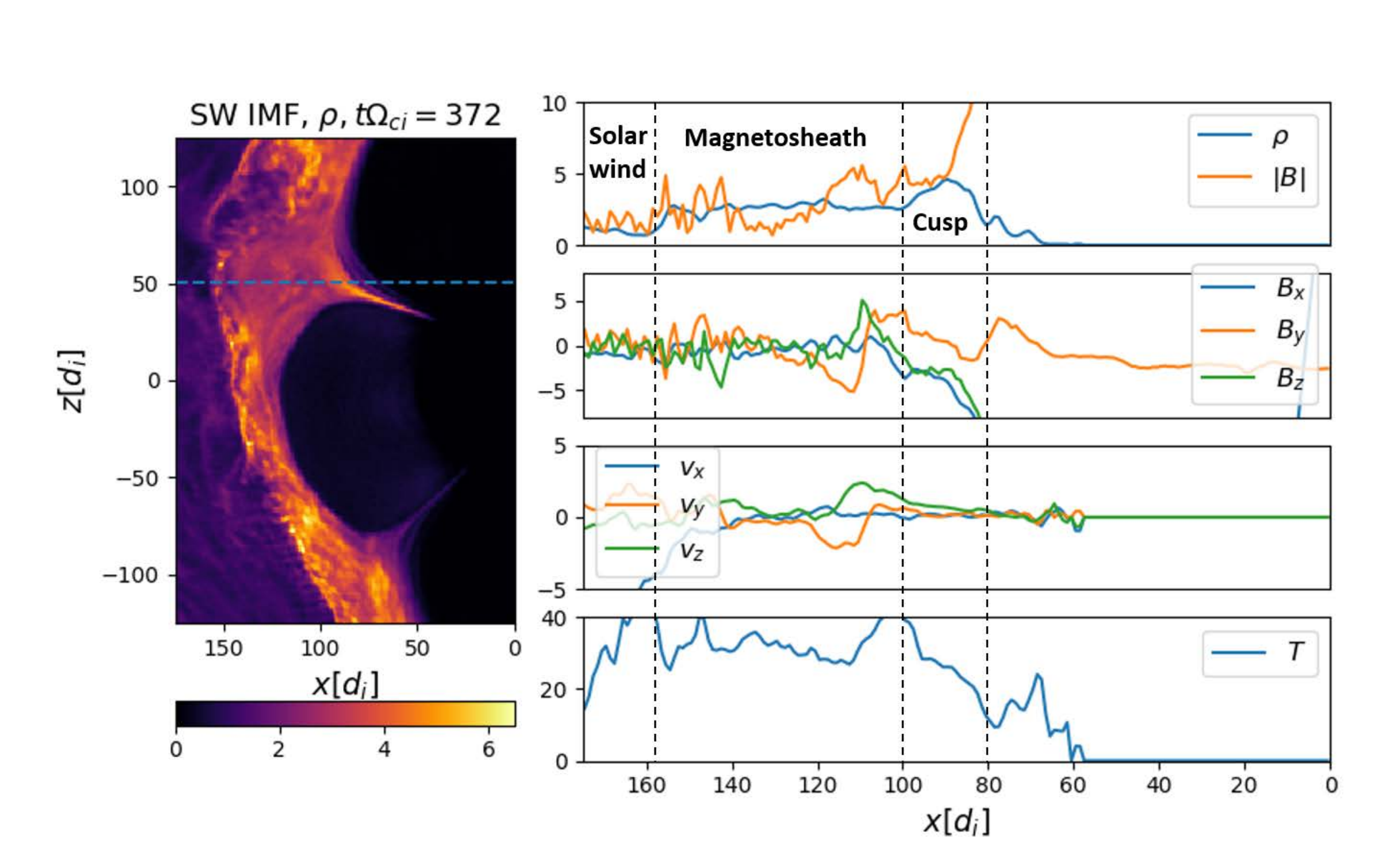}
  \caption{Left: Density in the southward IMF simulation at $t\Omega_{ci} = 372$. Dashed line shows the cut along which other quantities are computed. Right: Density, magnetic field, velocity and scalar temperature along the cut. }
  \label{fig:cut_sw2}
\end{figure}

In contrast, in Figs.~\ref{fig:panels_sw2} and \ref{fig:cut_sw2}, which correspond to a later time in the southward IMF case, there is no density enhancement in the magnetosheath close to the cusp, nor are there strong fluctuations in $v_z$. A clear positive $v_z$ signature can be seen at the magnetopause due to reconnection between the southward IMF and the dipole field (Fig.~\ref{fig:panels_sw2}). Compared to the earlier time in this simulation shown in Fig.~\ref{fig:panels_sw}, the density and magnetic field components do not show as much variation in the magnetosheath. Although there are still $B_z$ fluctuations in the magnetosheath, they do not occur in the region outside the cusp as at the earlier time ($t\Omega_{ci} = 281$), where oscillatory strong $\delta B_z$ can be seen in the top-right panel of Fig.~\ref{fig:fluctuations_sw}. 

The effect of the magnetosheath fluctuations on the cusp boundary can be seen at these two different times under southward IMF. As mentioned earlier, in \cite{lavraud:2005}, this boundary was associated with the region of strong velocity shear between the cusp and the magnetosheath. Because of the impact of the jet close to the cusp at $t\Omega_{ci} = 281$, the region with strong velocity shear at $z= 45$ is approximately $15$ $d_i$ further earthward than at $t\Omega_{ci} = 372$. The shape of the boundary is also more indented compared to the northward IMF case. 


\section{Discussion}

We now consider the observational context of these simulations. With respect to the global geometry of the system under northward and southward IMF conditions, the cusp positions and magnetopause locations show clear differences. The magnetopause position is further sunward in the northward IMF case, which is expected due to the erosion of the dayside magnetic field by magnetic reconnection for southward IMF. The cusp in the northward IMF simulation also appears at a higher latitude compared to the southward IMF case. This is consistent with observations, in which the cusp location is found to decrease in latitude for increasing southward IMF, while remaining insensitive to the magnitude of $B_z$ for northward IMF \cite{burch:1973,meng:1983,pitout:2006,pitout:2009}.

The global magnetic field geometry shows similarities to observations and MHD simulations of the magnetosphere under radial IMF conditions \cite{pi:2017,pi:2018}. However, compared to MHD simulations with radial IMF, the magnetosheath contains kinetic turbulent structures, which can be seen, for instance, by comparing the density profiles in this work and MHD studies \cite{pi:2018,tang:2013}. Under radial IMF with an earthward $B_x$ component, the magnetic field near the magnetopause is expected to be northward in the Northern Hemisphere, and southward in the Southern Hemisphere, with reconnection taking place tailward of the northern cusp and in the Southern Hemisphere at the dayside magnetopause. In  our simulations, the additional northward and southward IMF components slightly modify this picture. In the southward IMF case, there is indeed a quasi-steady reconnection site in the southern Hemisphere as seen from the kinked reconnected field lines in Fig.~\ref{fig:field_lines}. In the northward IMF case, there is no steady reconnection at the magnetopause, though the effect of turbulence on transient reconnection in the subsolar region  has been discussed \cite{chen:2021jet}. From Fig.~\ref{fig:field_lines}, the strongly kinked field lines associated with reconnection are found closer to the southern cusp, providing an upper $z$ limit for the location of the reconnection site. With respect to the reconnection tailward of the northern cusp, signatures are seen in both the field lines and flows for the northward IMF case, while they are not seen in the southward IMF case, though it is possible that the reconnection site could lie outside the simulation domain.



We now compare the characteristics of the exterior cusp and its magnetosheath boundary to the observations. In \citeA{lavraud:2002,lavraud:2004,lavraud:2005}, observational evidence of the exterior cusp  and its characteristics are presented. This region is diamagnetic and has boundaries with the lobes, dayside plasma sheet and the magnetosheath. Under northward IMF, the exterior cusp is stagnant, while under southward IMF, the region is convective. However, it should also be noted that these statistical studies  include a range of IMF directions while our simulations are done with quasi-radial IMF, so an exact correspondence should not be expected.

In the quasi-radial southward IMF case, the geometry is such that the quasi-parallel shock is close to the cusp, allowing foreshock turbulence and structures (e.g.~jets) to affect the magnetosheath immediately outside the cusp. We have shown the time-evolution of the magnetosheath and cusp after the impact of a jet, during which there are density enhancements both in the magnetosheath and the cusp. The boundary between the cusp and magnetosheath also becomes indented (compare the $|v_i|$ contours in Figs.~\ref{fig:panels_sw} and \ref{fig:panels_sw2}). This is consistent with observations which show indentations of the cusp boundary \cite{zhang:2007,pitout:2021}.


We now focus on the northward IMF case. In the observations, the perpendicular $x$ velocity in the cusp is small, with sunward convection at the poleward edge of the cusp. There are also downward flows, which may be indicative of reconnection tailward of the cusp \cite{lavraud:2002,lavraud:2004}. At the magnetosheath boundary, there is a strong velocity gradient, and a transition from super-Alfv\'enic to sub-Alfv\'enic flow \cite{lavraud:2005}. It has also been shown that a plasma depletion layer with reduced density and increased magnetic field can exist in this region \cite{lavraud:2004,zwan:1976,phan:2003,avanov:2001}. 

In the northward IMF simulation, there is a sharp boundary between the exterior cusp and the magnetosheath, as can be seen in Figs.~\ref{fig:panels_nw} and ~\ref{fig:cut_nw}. This boundary is not indented, as mentioned in the observations \cite{lavraud:2005}. There is a large velocity gradient, particularly in the $v_z$ component, and a small negative $v_z$ on the cusp side of the cusp-magnetosheath boundary ($x \approx 75$, $z\approx 75$). This is consistent with the observations and the presence of reconnection tailward of the cusp. In the simulation, while we find signatures of reconnection such as flux ropes and kinked field lines, the reconnection site is in the lower-resolution region outside the central domain and therefore its detailed analysis is not a focus of this study. The depletion layer is dynamic, with varying width and density over the course of the simulation, and the plasma is sub-Alfv\'enic in this layer. While there are flows in the cusp as shown in Fig.~\ref{fig:panels_nw}, it can be seen from the $|v_i|$ plot that aside from the abovementioned downward flow, their magnitude is small compared to the magnetosheath plasma. The characteristics of the depletion layer, such as the energy spectrum, show similarities to Cluster observations \cite{lavraud:2004}.

In the results of \citeA{lavraud:2002} for a cusp crossing under northward IMF conditions, the ion temperature shows a sharp decrease at the cusp-magnetosheath boundary. There is indeed a temperature decrease as shown in Fig.~\ref{fig:particle_nw}, but this is not the case at all times in the simulation. At other instances, temperature can also increase from the cusp to the magnetosheath. This difference could be attributed to the stronger radial component of the IMF in the simulation, and the heating of ions downstream of the quasi-parallel shock.

Finally, we discuss how our results relate to other hybrid simulation studies. In \citeA{grandin:2020},  purely southward and northward IMF simulations were performed using a two-dimensional hybrid-Vlasov model. These simulations show that flux-transfer events were correlated with proton precipitation at the cusp in the southward IMF case, while lobe reconnection contributes to precipitation in the northward IMF case. Our simulations focus on the turbulence under quasi-radial IMF and its  effects on the overall structure of the magnetosheath and cusp. In the southward IMF case, we show how turbulence and jets can lead to strong density enhancements in the magnetosheath and cause motion of the cusp boundary, which, in concert with reconnection, may cause density enhancements in the cusp. In contrast, the northward IMF case shows reconnection in regions tailward of the cusp. 
We note that while we do observe flux transfer events (FTEs) in the quasi-radial southward IMF case, these do not grow to the spatial scales seen in \citeA{grandin:2020}. This is likely because the quasi-radial IMF conditions are less favourable for flux transfer events \cite{wang:2006}.

\section{Summary}

To summarize, we have performed 3D global hybrid simulations  under northward and southward quasi-radial IMF. The southward quasi-radial IMF case shows stronger density fluctuations in the magnetosheath, particularly at the magnetopause boundary, while the northward IMF case has a pronounced diamagnetic region at the northern cusp, as well as a dynamic plasma depletion layer with reduced density and increased magnetic field. In the southward IMF case, foreshock turbulence leads to strong magnetosheath perturbations outside the cusp, including electromagnetic waves and high-speed jets. These cause motion of the cusp boundary, and together with magnetic reconnection at the subsolar magnetopause, contribute to the variation of density in the cusp.

The simulation results have also been compared to observations. In terms of the global structure, the relative positions of the cusp in the northward and southward IMF cases show agreement with observations. In the northward IMF case, a diamagnetic region with low ion flow velocity is found, which may be indicative of the stagnant exterior cusp as seen in observations. The presence of a sub-Alfv\'enic plasma depletion layer at the magnetosheath boundary is also consistent with theory and observations, though it appears more dynamic in the simulations due to the turbulent magnetosheath under quasi-radial IMF conditions. As mentioned earlier, we note that a limitation of this study is that the turbulence scales may be exaggerated due to the downscaled simulation size, though the kinetic physics of high-speed jets has been shown to be consistent with observations \cite{omelchenko:2021jet}.

\acknowledgments
This work was supported by NASA Grants 80NSSC21K1462, 80NSSC21K1046, 80NSSC20K1312, 80NSSC21K1483. The authors thank Johnny Chang, Michael Heinsohn, Nancy Carney, and other NASA Advanced-Supercomputing and High-End-Computing team members for their professional support to make the simulation possible. 

\section*{Open research}
The data that support the findings of this study are openly available at \cite{ng:2022hybriddata}. Movies supporting the paper are found at \cite{ng:2022movie}.


\bibliography{compare_rev_4}

\begin{thebibliography}{}

\bibitem [\protect \citeauthoryear {%
Avanov%
, Smirnov%
, Waite~Jr.%
, Fuselier%
\BCBL {}\ \BBA {} Vaisberg%
}{%
Avanov%
\ \protect \BOthers {.}}{%
{\protect \APACyear {2001}}%
}]{%
avanov:2001}
\APACinsertmetastar {%
avanov:2001}%
\begin{APACrefauthors}%
Avanov, L\BPBI A.%
, Smirnov, V\BPBI N.%
, Waite~Jr., J\BPBI H.%
, Fuselier, S\BPBI A.%
\BCBL {}\ \BBA {} Vaisberg, O\BPBI L.%
\end{APACrefauthors}%
\unskip\
\newblock
\APACrefYearMonthDay{2001}{}{}.
\newblock
{\BBOQ}\APACrefatitle {High-latitude magnetic reconnection in sub-Alfvénic
  flow: Interball Tail observations on May 29, 1996} {High-latitude magnetic
  reconnection in sub-alfvénic flow: Interball tail observations on may 29,
  1996}.{\BBCQ}
\newblock
\APACjournalVolNumPages{Journal of Geophysical Research: Space
  Physics}{106}{A12}{29491-29502}.
\newblock
\begin{APACrefURL}
  \url{https://agupubs.onlinelibrary.wiley.com/doi/abs/10.1029/2000JA000460}
  \end{APACrefURL}
\newblock
\begin{APACrefDOI} \doi{https://doi.org/10.1029/2000JA000460} \end{APACrefDOI}
\PrintBackRefs{\CurrentBib}

\bibitem [\protect \citeauthoryear {%
Blanco-Cano%
, Omidi%
\BCBL {}\ \BBA {} Russell%
}{%
Blanco-Cano%
\ \protect \BOthers {.}}{%
{\protect \APACyear {2009}}%
}]{%
blancocano:2009}
\APACinsertmetastar {%
blancocano:2009}%
\begin{APACrefauthors}%
Blanco-Cano, X.%
, Omidi, N.%
\BCBL {}\ \BBA {} Russell, C\BPBI T.%
\end{APACrefauthors}%
\unskip\
\newblock
\APACrefYearMonthDay{2009}{}{}.
\newblock
{\BBOQ}\APACrefatitle {Global hybrid simulations: Foreshock waves and cavitons
  under radial interplanetary magnetic field geometry} {Global hybrid
  simulations: Foreshock waves and cavitons under radial interplanetary
  magnetic field geometry}.{\BBCQ}
\newblock
\APACjournalVolNumPages{Journal of Geophysical Research: Space
  Physics}{114}{A1}{}.
\newblock
\begin{APACrefURL}
  \url{https://agupubs.onlinelibrary.wiley.com/doi/abs/10.1029/2008JA013406}
  \end{APACrefURL}
\newblock
\begin{APACrefDOI} \doi{https://doi.org/10.1029/2008JA013406} \end{APACrefDOI}
\PrintBackRefs{\CurrentBib}

\bibitem [\protect \citeauthoryear {%
Burch%
}{%
Burch%
}{%
{\protect \APACyear {1973}}%
}]{%
burch:1973}
\APACinsertmetastar {%
burch:1973}%
\begin{APACrefauthors}%
Burch, J\BPBI L.%
\end{APACrefauthors}%
\unskip\
\newblock
\APACrefYearMonthDay{1973}{}{}.
\newblock
{\BBOQ}\APACrefatitle {Rate of erosion of dayside magnetic flux based on a
  quantitative study of the dependence of polar cusp latitude on the
  interplanetary magnetic field} {Rate of erosion of dayside magnetic flux
  based on a quantitative study of the dependence of polar cusp latitude on the
  interplanetary magnetic field}.{\BBCQ}
\newblock
\APACjournalVolNumPages{Radio Science}{8}{11}{955-961}.
\newblock
\begin{APACrefURL}
  \url{https://agupubs.onlinelibrary.wiley.com/doi/abs/10.1029/RS008i011p00955}
  \end{APACrefURL}
\newblock
\begin{APACrefDOI} \doi{https://doi.org/10.1029/RS008i011p00955}
  \end{APACrefDOI}
\PrintBackRefs{\CurrentBib}

\bibitem [\protect \citeauthoryear {%
Burkholder%
, Nykyri%
\BCBL {}\ \BBA {} Ma%
}{%
Burkholder%
\ \protect \BOthers {.}}{%
{\protect \APACyear {2021}}%
}]{%
burkholder:2021}
\APACinsertmetastar {%
burkholder:2021}%
\begin{APACrefauthors}%
Burkholder, B\BPBI L.%
, Nykyri, K.%
\BCBL {}\ \BBA {} Ma, X.%
\end{APACrefauthors}%
\unskip\
\newblock
\APACrefYearMonthDay{2021}{}{}.
\newblock
{\BBOQ}\APACrefatitle {Magnetospheric Multiscale Statistics of High Energy
  Electrons Trapped in Diamagnetic Cavities} {Magnetospheric multiscale
  statistics of high energy electrons trapped in diamagnetic cavities}.{\BBCQ}
\newblock
\APACjournalVolNumPages{Journal of Geophysical Research: Space
  Physics}{126}{1}{e2020JA028341}.
\newblock
\begin{APACrefURL}
  \url{https://agupubs.onlinelibrary.wiley.com/doi/abs/10.1029/2020JA028341}
  \end{APACrefURL}
\newblock
\APACrefnote{e2020JA028341 2020JA028341}
\newblock
\begin{APACrefDOI} \doi{https://doi.org/10.1029/2020JA028341} \end{APACrefDOI}
\PrintBackRefs{\CurrentBib}

\bibitem [\protect \citeauthoryear {%
Chen%
, Ng%
, Omelchenko%
\BCBL {}\ \BBA {} Wang%
}{%
Chen%
, Ng%
\BCBL {}\ \protect \BOthers {.}}{%
{\protect \APACyear {2021}}%
}]{%
chen:2021jet}
\APACinsertmetastar {%
chen:2021jet}%
\begin{APACrefauthors}%
Chen, L\BHBI J.%
, Ng, J.%
, Omelchenko, Y.%
\BCBL {}\ \BBA {} Wang, S.%
\end{APACrefauthors}%
\unskip\
\newblock
\APACrefYearMonthDay{2021}{}{}.
\newblock
{\BBOQ}\APACrefatitle {Magnetopause Reconnection and Indents Induced by
  Foreshock Turbulence} {Magnetopause reconnection and indents induced by
  foreshock turbulence}.{\BBCQ}
\newblock
\APACjournalVolNumPages{Geophysical Research Letters}{48}{11}{e2021GL093029}.
\newblock
\begin{APACrefURL}
  \url{https://agupubs.onlinelibrary.wiley.com/doi/abs/10.1029/2021GL093029}
  \end{APACrefURL}
\newblock
\begin{APACrefDOI} \doi{https://doi.org/10.1029/2021GL093029} \end{APACrefDOI}
\PrintBackRefs{\CurrentBib}

\bibitem [\protect \citeauthoryear {%
Chen%
, Wang%
\BCBL {}\ \protect \BOthers {.}}{%
Chen%
, Wang%
\BCBL {}\ \protect \BOthers {.}}{%
{\protect \APACyear {2021}}%
}]{%
chen:2021}
\APACinsertmetastar {%
chen:2021}%
\begin{APACrefauthors}%
Chen, L\BHBI J.%
, Wang, S.%
, Ng, J.%
, Bessho, N.%
, Tang, J\BHBI M.%
, Fung, S\BPBI F.%
\BDBL {}Burch, J.%
\end{APACrefauthors}%
\unskip\
\newblock
\APACrefYearMonthDay{2021}{}{}.
\newblock
{\BBOQ}\APACrefatitle {Solitary Magnetic Structures at Quasi-Parallel
  Collisionless Shocks: Formation} {Solitary magnetic structures at
  quasi-parallel collisionless shocks: Formation}.{\BBCQ}
\newblock
\APACjournalVolNumPages{Geophysical Research Letters}{48}{1}{e2020GL090800}.
\newblock
\begin{APACrefURL}
  \url{https://agupubs.onlinelibrary.wiley.com/doi/abs/10.1029/2020GL090800}
  \end{APACrefURL}
\newblock
\APACrefnote{e2020GL090800 2020GL090800}
\newblock
\begin{APACrefDOI} \doi{https://doi.org/10.1029/2020GL090800} \end{APACrefDOI}
\PrintBackRefs{\CurrentBib}

\bibitem [\protect \citeauthoryear {%
Delcourt%
\ \BBA {} Sauvaud%
}{%
Delcourt%
\ \BBA {} Sauvaud%
}{%
{\protect \APACyear {1999}}%
}]{%
delcourt:1999}
\APACinsertmetastar {%
delcourt:1999}%
\begin{APACrefauthors}%
Delcourt, D\BPBI C.%
\BCBT {}\ \BBA {} Sauvaud, J\BHBI A.%
\end{APACrefauthors}%
\unskip\
\newblock
\APACrefYearMonthDay{1999}{}{}.
\newblock
{\BBOQ}\APACrefatitle {Populating of the cusp and boundary layers by energetic
  (hundreds of keV) equatorial particles} {Populating of the cusp and boundary
  layers by energetic (hundreds of kev) equatorial particles}.{\BBCQ}
\newblock
\APACjournalVolNumPages{Journal of Geophysical Research: Space
  Physics}{104}{A10}{22635-22648}.
\newblock
\begin{APACrefURL}
  \url{https://agupubs.onlinelibrary.wiley.com/doi/abs/10.1029/1999JA900251}
  \end{APACrefURL}
\newblock
\begin{APACrefDOI} \doi{https://doi.org/10.1029/1999JA900251} \end{APACrefDOI}
\PrintBackRefs{\CurrentBib}

\bibitem [\protect \citeauthoryear {%
Frank%
}{%
Frank%
}{%
{\protect \APACyear {1971}}%
}]{%
frank:1971}
\APACinsertmetastar {%
frank:1971}%
\begin{APACrefauthors}%
Frank, L\BPBI A.%
\end{APACrefauthors}%
\unskip\
\newblock
\APACrefYearMonthDay{1971}{}{}.
\newblock
{\BBOQ}\APACrefatitle {Plasma in the Earth's polar magnetosphere} {Plasma in
  the earth's polar magnetosphere}.{\BBCQ}
\newblock
\APACjournalVolNumPages{Journal of Geophysical Research
  (1896-1977)}{76}{22}{5202-5219}.
\newblock
\begin{APACrefURL}
  \url{https://agupubs.onlinelibrary.wiley.com/doi/abs/10.1029/JA076i022p05202}
  \end{APACrefURL}
\newblock
\begin{APACrefDOI} \doi{https://doi.org/10.1029/JA076i022p05202}
  \end{APACrefDOI}
\PrintBackRefs{\CurrentBib}

\bibitem [\protect \citeauthoryear {%
Gary%
}{%
Gary%
}{%
{\protect \APACyear {1991}}%
}]{%
gary:1991}
\APACinsertmetastar {%
gary:1991}%
\begin{APACrefauthors}%
Gary, S\BPBI P.%
\end{APACrefauthors}%
\unskip\
\newblock
\APACrefYearMonthDay{1991}{}{}.
\newblock
{\BBOQ}\APACrefatitle {Electromagnetic ion/ion instabilities and their
  consequences in space plasmas: A review} {Electromagnetic ion/ion
  instabilities and their consequences in space plasmas: A review}.{\BBCQ}
\newblock
\APACjournalVolNumPages{Space Science Reviews}{56}{3-4}{373--415}.
\PrintBackRefs{\CurrentBib}

\bibitem [\protect \citeauthoryear {%
Grandin%
\ \protect \BOthers {.}}{%
Grandin%
\ \protect \BOthers {.}}{%
{\protect \APACyear {2020}}%
}]{%
grandin:2020}
\APACinsertmetastar {%
grandin:2020}%
\begin{APACrefauthors}%
Grandin, M.%
, Turc, L.%
, Battarbee, M.%
, Ganse, U.%
, Johlander, A.%
, Pfau-Kempf, Y.%
\BDBL {}Palmroth, M.%
\end{APACrefauthors}%
\unskip\
\newblock
\APACrefYearMonthDay{2020}{}{}.
\newblock
{\BBOQ}\APACrefatitle {Hybrid-Vlasov simulation of auroral proton precipitation
  in the cusps: Comparison of northward and southward interplanetary magnetic
  field driving} {Hybrid-vlasov simulation of auroral proton precipitation in
  the cusps: Comparison of northward and southward interplanetary magnetic
  field driving}.{\BBCQ}
\newblock
\APACjournalVolNumPages{Journal of Space Weather and Space Climate}{}{}{}.
\PrintBackRefs{\CurrentBib}

\bibitem [\protect \citeauthoryear {%
Heikkila%
\ \BBA {} Winningham%
}{%
Heikkila%
\ \BBA {} Winningham%
}{%
{\protect \APACyear {1971}}%
}]{%
heikkila:1971}
\APACinsertmetastar {%
heikkila:1971}%
\begin{APACrefauthors}%
Heikkila, W\BPBI J.%
\BCBT {}\ \BBA {} Winningham, J\BPBI D.%
\end{APACrefauthors}%
\unskip\
\newblock
\APACrefYearMonthDay{1971}{}{}.
\newblock
{\BBOQ}\APACrefatitle {Penetration of magnetosheath plasma to low altitudes
  through the dayside magnetospheric cusps} {Penetration of magnetosheath
  plasma to low altitudes through the dayside magnetospheric cusps}.{\BBCQ}
\newblock
\APACjournalVolNumPages{Journal of Geophysical Research
  (1896-1977)}{76}{4}{883-891}.
\newblock
\begin{APACrefURL}
  \url{https://agupubs.onlinelibrary.wiley.com/doi/abs/10.1029/JA076i004p00883}
  \end{APACrefURL}
\newblock
\begin{APACrefDOI} \doi{https://doi.org/10.1029/JA076i004p00883}
  \end{APACrefDOI}
\PrintBackRefs{\CurrentBib}

\bibitem [\protect \citeauthoryear {%
Hietala%
\ \protect \BOthers {.}}{%
Hietala%
\ \protect \BOthers {.}}{%
{\protect \APACyear {2012}}%
}]{%
hietala:2012}
\APACinsertmetastar {%
hietala:2012}%
\begin{APACrefauthors}%
Hietala, H.%
, Partamies, N.%
, Laitinen, T.%
, Clausen, L\BPBI B.%
, Facsk{\'o}, G.%
, Vaivads, A.%
\BDBL {}others%
\end{APACrefauthors}%
\unskip\
\newblock
\APACrefYearMonthDay{2012}{}{}.
\newblock
{\BBOQ}\APACrefatitle {Supermagnetosonic subsolar magnetosheath jets and their
  effects: from the solar wind to the ionospheric convection}
  {Supermagnetosonic subsolar magnetosheath jets and their effects: from the
  solar wind to the ionospheric convection}.{\BBCQ}
\newblock
\BIn{} \APACrefbtitle {Annales Geophysicae.} {Annales geophysicae.}
\PrintBackRefs{\CurrentBib}

\bibitem [\protect \citeauthoryear {%
Kajdi{\v{c}}%
, Blanco-Cano%
, Omidi%
\BCBL {}\ \BBA {} Russell%
}{%
Kajdi{\v{c}}%
\ \protect \BOthers {.}}{%
{\protect \APACyear {2011}}%
}]{%
kajdivc:2011}
\APACinsertmetastar {%
kajdivc:2011}%
\begin{APACrefauthors}%
Kajdi{\v{c}}, P.%
, Blanco-Cano, X.%
, Omidi, N.%
\BCBL {}\ \BBA {} Russell, C\BPBI T.%
\end{APACrefauthors}%
\unskip\
\newblock
\APACrefYearMonthDay{2011}{}{}.
\newblock
{\BBOQ}\APACrefatitle {Multi-spacecraft study of foreshock cavitons upstream of
  the quasi-parallel bow shock} {Multi-spacecraft study of foreshock cavitons
  upstream of the quasi-parallel bow shock}.{\BBCQ}
\newblock
\APACjournalVolNumPages{Planetary and Space Science}{59}{8}{705--714}.
\PrintBackRefs{\CurrentBib}

\bibitem [\protect \citeauthoryear {%
Karimabadi%
\ \protect \BOthers {.}}{%
Karimabadi%
\ \protect \BOthers {.}}{%
{\protect \APACyear {2014}}%
}]{%
karimabadi:2014}
\APACinsertmetastar {%
karimabadi:2014}%
\begin{APACrefauthors}%
Karimabadi, H.%
, Roytershteyn, V.%
, Vu, H\BPBI X.%
, Omelchenko, Y\BPBI A.%
, Scudder, J.%
, Daughton, W.%
\BDBL {}Geveci, B.%
\end{APACrefauthors}%
\unskip\
\newblock
\APACrefYearMonthDay{2014}{}{}.
\newblock
{\BBOQ}\APACrefatitle {The link between shocks, turbulence, and magnetic
  reconnection in collisionless plasmas} {The link between shocks, turbulence,
  and magnetic reconnection in collisionless plasmas}.{\BBCQ}
\newblock
\APACjournalVolNumPages{Physics of Plasmas}{21}{6}{062308}.
\newblock
\begin{APACrefURL} \url{https://doi.org/10.1063/1.4882875} \end{APACrefURL}
\newblock
\begin{APACrefDOI} \doi{10.1063/1.4882875} \end{APACrefDOI}
\PrintBackRefs{\CurrentBib}

\bibitem [\protect \citeauthoryear {%
Kempf%
\ \protect \BOthers {.}}{%
Kempf%
\ \protect \BOthers {.}}{%
{\protect \APACyear {2015}}%
}]{%
kempf:2015}
\APACinsertmetastar {%
kempf:2015}%
\begin{APACrefauthors}%
Kempf, Y.%
, Pokhotelov, D.%
, Gutynska, O.%
, Wilson~III, L\BPBI B.%
, Walsh, B\BPBI M.%
, Alfthan, S\BPBI v.%
\BDBL {}Palmroth, M.%
\end{APACrefauthors}%
\unskip\
\newblock
\APACrefYearMonthDay{2015}{}{}.
\newblock
{\BBOQ}\APACrefatitle {Ion distributions in the Earth's foreshock:
  Hybrid-Vlasov simulation and THEMIS observations} {Ion distributions in the
  earth's foreshock: Hybrid-vlasov simulation and themis observations}.{\BBCQ}
\newblock
\APACjournalVolNumPages{Journal of Geophysical Research: Space
  Physics}{120}{5}{3684-3701}.
\newblock
\begin{APACrefURL}
  \url{https://agupubs.onlinelibrary.wiley.com/doi/abs/10.1002/2014JA020519}
  \end{APACrefURL}
\newblock
\begin{APACrefDOI} \doi{https://doi.org/10.1002/2014JA020519} \end{APACrefDOI}
\PrintBackRefs{\CurrentBib}

\bibitem [\protect \citeauthoryear {%
Lavraud%
\ \protect \BOthers {.}}{%
Lavraud%
\ \protect \BOthers {.}}{%
{\protect \APACyear {2002}}%
}]{%
lavraud:2002}
\APACinsertmetastar {%
lavraud:2002}%
\begin{APACrefauthors}%
Lavraud, B.%
, Dunlop, M\BPBI W.%
, Phan, T\BPBI D.%
, Rème, H.%
, Bosqued, J\BHBI M.%
, Dandouras, I.%
\BDBL {}Balogh, A.%
\end{APACrefauthors}%
\unskip\
\newblock
\APACrefYearMonthDay{2002}{}{}.
\newblock
{\BBOQ}\APACrefatitle {Cluster observations of the exterior cusp and its
  surrounding boundaries under northward IMF} {Cluster observations of the
  exterior cusp and its surrounding boundaries under northward imf}.{\BBCQ}
\newblock
\APACjournalVolNumPages{Geophysical Research Letters}{29}{20}{56-1-56-4}.
\newblock
\begin{APACrefURL}
  \url{https://agupubs.onlinelibrary.wiley.com/doi/abs/10.1029/2002GL015464}
  \end{APACrefURL}
\newblock
\begin{APACrefDOI} \doi{https://doi.org/10.1029/2002GL015464} \end{APACrefDOI}
\PrintBackRefs{\CurrentBib}

\bibitem [\protect \citeauthoryear {%
Lavraud%
\ \protect \BOthers {.}}{%
Lavraud%
\ \protect \BOthers {.}}{%
{\protect \APACyear {2005}}%
}]{%
lavraud:2005}
\APACinsertmetastar {%
lavraud:2005}%
\begin{APACrefauthors}%
Lavraud, B.%
, Fedorov, A.%
, Budnik, E.%
, Thomsen, M\BPBI F.%
, Grigoriev, A.%
, Cargill, P\BPBI J.%
\BDBL {}Balogh, A.%
\end{APACrefauthors}%
\unskip\
\newblock
\APACrefYearMonthDay{2005}{}{}.
\newblock
{\BBOQ}\APACrefatitle {High-altitude cusp flow dependence on IMF orientation: A
  3-year Cluster statistical study} {High-altitude cusp flow dependence on imf
  orientation: A 3-year cluster statistical study}.{\BBCQ}
\newblock
\APACjournalVolNumPages{Journal of Geophysical Research: Space
  Physics}{110}{A2}{}.
\newblock
\begin{APACrefURL}
  \url{https://agupubs.onlinelibrary.wiley.com/doi/abs/10.1029/2004JA010804}
  \end{APACrefURL}
\newblock
\begin{APACrefDOI} \doi{https://doi.org/10.1029/2004JA010804} \end{APACrefDOI}
\PrintBackRefs{\CurrentBib}

\bibitem [\protect \citeauthoryear {%
Lavraud%
\ \protect \BOthers {.}}{%
Lavraud%
\ \protect \BOthers {.}}{%
{\protect \APACyear {2004}}%
}]{%
lavraud:2004}
\APACinsertmetastar {%
lavraud:2004}%
\begin{APACrefauthors}%
Lavraud, B.%
, Phan, T\BPBI D.%
, Dunlop, M\BPBI W.%
, Taylor, M\BPBI G\BPBI G\BPBI G\BPBI T.%
, Cargill, P\BPBI J.%
, Bosqued, J\BHBI M.%
\BDBL {}Fazakerley, A.%
\end{APACrefauthors}%
\unskip\
\newblock
\APACrefYearMonthDay{2004}{}{}.
\newblock
{\BBOQ}\APACrefatitle {The exterior cusp and its boundary with the
  magnetosheath: Cluster multi-event analysis} {The exterior cusp and its
  boundary with the magnetosheath: Cluster multi-event analysis}.{\BBCQ}
\newblock
\APACjournalVolNumPages{Annales Geophysicae}{22}{8}{3039--3054}.
\newblock
\begin{APACrefURL} \url{https://angeo.copernicus.org/articles/22/3039/2004/}
  \end{APACrefURL}
\newblock
\begin{APACrefDOI} \doi{10.5194/angeo-22-3039-2004} \end{APACrefDOI}
\PrintBackRefs{\CurrentBib}

\bibitem [\protect \citeauthoryear {%
Lin%
\ \BBA {} Wang%
}{%
Lin%
\ \BBA {} Wang%
}{%
{\protect \APACyear {2005}}%
}]{%
lin:2005}
\APACinsertmetastar {%
lin:2005}%
\begin{APACrefauthors}%
Lin, Y.%
\BCBT {}\ \BBA {} Wang, X\BPBI Y.%
\end{APACrefauthors}%
\unskip\
\newblock
\APACrefYearMonthDay{2005}{}{}.
\newblock
{\BBOQ}\APACrefatitle {Three-dimensional global hybrid simulation of dayside
  dynamics associated with the quasi-parallel bow shock} {Three-dimensional
  global hybrid simulation of dayside dynamics associated with the
  quasi-parallel bow shock}.{\BBCQ}
\newblock
\APACjournalVolNumPages{Journal of Geophysical Research: Space
  Physics}{110}{A12}{}.
\newblock
\begin{APACrefURL}
  \url{https://agupubs.onlinelibrary.wiley.com/doi/abs/10.1029/2005JA011243}
  \end{APACrefURL}
\newblock
\begin{APACrefDOI} \doi{https://doi.org/10.1029/2005JA011243} \end{APACrefDOI}
\PrintBackRefs{\CurrentBib}

\bibitem [\protect \citeauthoryear {%
Meng%
}{%
Meng%
}{%
{\protect \APACyear {1983}}%
}]{%
meng:1983}
\APACinsertmetastar {%
meng:1983}%
\begin{APACrefauthors}%
Meng, C\BHBI I.%
\end{APACrefauthors}%
\unskip\
\newblock
\APACrefYearMonthDay{1983}{}{}.
\newblock
{\BBOQ}\APACrefatitle {Case studies of the storm time variation of the polar
  cusp} {Case studies of the storm time variation of the polar cusp}.{\BBCQ}
\newblock
\APACjournalVolNumPages{Journal of Geophysical Research: Space
  Physics}{88}{A1}{137-149}.
\newblock
\begin{APACrefURL}
  \url{https://agupubs.onlinelibrary.wiley.com/doi/abs/10.1029/JA088iA01p00137}
  \end{APACrefURL}
\newblock
\begin{APACrefDOI} \doi{https://doi.org/10.1029/JA088iA01p00137}
  \end{APACrefDOI}
\PrintBackRefs{\CurrentBib}

\bibitem [\protect \citeauthoryear {%
Ng%
, Chen%
, Omelchenko%
, Zou%
\BCBL {}\ \BBA {} Lavraud%
}{%
Ng%
\ \protect \BOthers {.}}{%
{\protect \APACyear {2022}}%
{\protect \APACexlab {{\protect \BCnt {1}}}}}]{%
ng:2022hybriddata}
\APACinsertmetastar {%
ng:2022hybriddata}%
\begin{APACrefauthors}%
Ng, J.%
, Chen, L\BHBI J.%
, Omelchenko, Y.%
, Zou, Y.%
\BCBL {}\ \BBA {} Lavraud, B.%
\end{APACrefauthors}%
\unskip\
\newblock
\APACrefYearMonthDay{2022{\protect \BCnt {1}}}{{\APACmonth{01}}}{}.
\newblock
\APACrefbtitle {{Dataset for "Hybrid simulations of the cusp and dayside
  magnetosheath dynamics under quasi-radial interplanetary magnetic fields"}.}
  {{Dataset for "Hybrid simulations of the cusp and dayside magnetosheath
  dynamics under quasi-radial interplanetary magnetic fields"}.}
\newblock
\APACaddressPublisher{}{Zenodo}.
\newblock
\begin{APACrefURL} \url{https://doi.org/10.5281/zenodo.5889594}
  \end{APACrefURL}
\newblock
\begin{APACrefDOI} \doi{10.5281/zenodo.5889594} \end{APACrefDOI}
\PrintBackRefs{\CurrentBib}

\bibitem [\protect \citeauthoryear {%
Ng%
, Chen%
, Omelchenko%
, Zou%
\BCBL {}\ \BBA {} Lavraud%
}{%
Ng%
\ \protect \BOthers {.}}{%
{\protect \APACyear {2022}}%
{\protect \APACexlab {{\protect \BCnt {2}}}}}]{%
ng:2022movie}
\APACinsertmetastar {%
ng:2022movie}%
\begin{APACrefauthors}%
Ng, J.%
, Chen, L\BHBI J.%
, Omelchenko, Y.%
, Zou, Y.%
\BCBL {}\ \BBA {} Lavraud, B.%
\end{APACrefauthors}%
\unskip\
\newblock
\APACrefYearMonthDay{2022{\protect \BCnt {2}}}{{\APACmonth{08}}}{}.
\newblock
\APACrefbtitle {{Movies for "Hybrid simulations of the cusp and dayside
  magnetosheath dynamics under quasi-radial interplanetary magnetic fields"}.}
  {{Movies for "Hybrid simulations of the cusp and dayside magnetosheath
  dynamics under quasi-radial interplanetary magnetic fields"}.}
\newblock
\APACaddressPublisher{}{Zenodo}.
\newblock
\begin{APACrefURL} \url{https://doi.org/10.5281/zenodo.6986351}
  \end{APACrefURL}
\newblock
\begin{APACrefDOI} \doi{10.5281/zenodo.6986351} \end{APACrefDOI}
\PrintBackRefs{\CurrentBib}

\bibitem [\protect \citeauthoryear {%
Ng%
, Chen%
\BCBL {}\ \BBA {} Omelchenko%
}{%
Ng%
\ \protect \BOthers {.}}{%
{\protect \APACyear {2021}}%
}]{%
ng:2021}
\APACinsertmetastar {%
ng:2021}%
\begin{APACrefauthors}%
Ng, J.%
, Chen, L\BHBI J.%
\BCBL {}\ \BBA {} Omelchenko, Y\BPBI A.%
\end{APACrefauthors}%
\unskip\
\newblock
\APACrefYearMonthDay{2021}{}{}.
\newblock
{\BBOQ}\APACrefatitle {Bursty magnetic reconnection at the Earth's magnetopause
  triggered by high-speed jets} {Bursty magnetic reconnection at the earth's
  magnetopause triggered by high-speed jets}.{\BBCQ}
\newblock
\APACjournalVolNumPages{Physics of Plasmas}{28}{9}{092902}.
\newblock
\begin{APACrefURL} \url{https://doi.org/10.1063/5.0054394} \end{APACrefURL}
\newblock
\begin{APACrefDOI} \doi{10.1063/5.0054394} \end{APACrefDOI}
\PrintBackRefs{\CurrentBib}

\bibitem [\protect \citeauthoryear {%
Nykyri%
, Otto%
, Adamson%
, Dougal%
\BCBL {}\ \BBA {} Mumme%
}{%
Nykyri%
\ \protect \BOthers {.}}{%
{\protect \APACyear {2011}}%
}]{%
nykyri:2011}
\APACinsertmetastar {%
nykyri:2011}%
\begin{APACrefauthors}%
Nykyri, K.%
, Otto, A.%
, Adamson, E.%
, Dougal, E.%
\BCBL {}\ \BBA {} Mumme, J.%
\end{APACrefauthors}%
\unskip\
\newblock
\APACrefYearMonthDay{2011}{}{}.
\newblock
{\BBOQ}\APACrefatitle {Cluster observations of a cusp diamagnetic cavity:
  Structure, size, and dynamics} {Cluster observations of a cusp diamagnetic
  cavity: Structure, size, and dynamics}.{\BBCQ}
\newblock
\APACjournalVolNumPages{Journal of Geophysical Research: Space
  Physics}{116}{A3}{}.
\newblock
\begin{APACrefURL}
  \url{https://agupubs.onlinelibrary.wiley.com/doi/abs/10.1029/2010JA015897}
  \end{APACrefURL}
\newblock
\begin{APACrefDOI} \doi{https://doi.org/10.1029/2010JA015897} \end{APACrefDOI}
\PrintBackRefs{\CurrentBib}

\bibitem [\protect \citeauthoryear {%
Nykyri%
, Otto%
, Adamson%
, Kronberg%
\BCBL {}\ \BBA {} Daly%
}{%
Nykyri%
\ \protect \BOthers {.}}{%
{\protect \APACyear {2012}}%
}]{%
nykyri:2012}
\APACinsertmetastar {%
nykyri:2012}%
\begin{APACrefauthors}%
Nykyri, K.%
, Otto, A.%
, Adamson, E.%
, Kronberg, E.%
\BCBL {}\ \BBA {} Daly, P.%
\end{APACrefauthors}%
\unskip\
\newblock
\APACrefYearMonthDay{2012}{}{}.
\newblock
{\BBOQ}\APACrefatitle {On the origin of high-energy particles in the cusp
  diamagnetic cavity} {On the origin of high-energy particles in the cusp
  diamagnetic cavity}.{\BBCQ}
\newblock
\APACjournalVolNumPages{Journal of Atmospheric and Solar-Terrestrial
  Physics}{87-88}{}{70-81}.
\newblock
\begin{APACrefURL}
  \url{https://www.sciencedirect.com/science/article/pii/S1364682611002495}
  \end{APACrefURL}
\newblock
\APACrefnote{Physical Process in the Cusp: Plasma Transport and Energization}
\newblock
\begin{APACrefDOI} \doi{https://doi.org/10.1016/j.jastp.2011.08.012}
  \end{APACrefDOI}
\PrintBackRefs{\CurrentBib}

\bibitem [\protect \citeauthoryear {%
Y.~Omelchenko%
\ \BBA {} Karimabadi%
}{%
Y.~Omelchenko%
\ \BBA {} Karimabadi%
}{%
{\protect \APACyear {2012}}%
}]{%
omelchenko:2012}
\APACinsertmetastar {%
omelchenko:2012}%
\begin{APACrefauthors}%
Omelchenko, Y.%
\BCBT {}\ \BBA {} Karimabadi, H.%
\end{APACrefauthors}%
\unskip\
\newblock
\APACrefYearMonthDay{2012}{}{}.
\newblock
{\BBOQ}\APACrefatitle {HYPERS: A unidimensional asynchronous framework for
  multiscale hybrid simulations} {Hypers: A unidimensional asynchronous
  framework for multiscale hybrid simulations}.{\BBCQ}
\newblock
\APACjournalVolNumPages{Journal of Computational Physics}{231}{4}{1766 - 1780}.
\newblock
\begin{APACrefURL}
  \url{http://www.sciencedirect.com/science/article/pii/S0021999111006462}
  \end{APACrefURL}
\newblock
\begin{APACrefDOI} \doi{https://doi.org/10.1016/j.jcp.2011.11.004}
  \end{APACrefDOI}
\PrintBackRefs{\CurrentBib}

\bibitem [\protect \citeauthoryear {%
Y\BPBI A.~Omelchenko%
}{%
Y\BPBI A.~Omelchenko%
}{%
{\protect \APACyear {2015}}%
}]{%
omelchenko:2015}
\APACinsertmetastar {%
omelchenko:2015}%
\begin{APACrefauthors}%
Omelchenko, Y\BPBI A.%
\end{APACrefauthors}%
\unskip\
\newblock
\APACrefYearMonthDay{2015}{Aug}{}.
\newblock
{\BBOQ}\APACrefatitle {Formation, spin-up, and stability of field-reversed
  configurations} {Formation, spin-up, and stability of field-reversed
  configurations}.{\BBCQ}
\newblock
\APACjournalVolNumPages{Phys. Rev. E}{92}{}{023105}.
\newblock
\begin{APACrefURL} \url{https://link.aps.org/doi/10.1103/PhysRevE.92.023105}
  \end{APACrefURL}
\newblock
\begin{APACrefDOI} \doi{10.1103/PhysRevE.92.023105} \end{APACrefDOI}
\PrintBackRefs{\CurrentBib}

\bibitem [\protect \citeauthoryear {%
Y\BPBI A.~Omelchenko%
, Chen%
\BCBL {}\ \BBA {} Ng%
}{%
Y\BPBI A.~Omelchenko%
\ \protect \BOthers {.}}{%
{\protect \APACyear {2021a}}%
}]{%
omelchenko:2021jet}
\APACinsertmetastar {%
omelchenko:2021jet}%
\begin{APACrefauthors}%
Omelchenko, Y\BPBI A.%
, Chen, L\BHBI J.%
\BCBL {}\ \BBA {} Ng, J.%
\end{APACrefauthors}%
\unskip\
\newblock
\APACrefYearMonthDay{2021a}{}{}.
\newblock
{\BBOQ}\APACrefatitle {3D Space-Time Adaptive Hybrid Simulations of
  Magnetosheath High-Speed Jets} {3d space-time adaptive hybrid simulations of
  magnetosheath high-speed jets}.{\BBCQ}
\newblock
\APACjournalVolNumPages{Journal of Geophysical Research: Space
  Physics}{126}{7}{e2020JA029035}.
\newblock
\begin{APACrefURL}
  \url{https://agupubs.onlinelibrary.wiley.com/doi/abs/10.1029/2020JA029035}
  \end{APACrefURL}
\newblock
\begin{APACrefDOI} \doi{https://doi.org/10.1029/2020JA029035} \end{APACrefDOI}
\PrintBackRefs{\CurrentBib}

\bibitem [\protect \citeauthoryear {%
Y\BPBI A.~Omelchenko%
, Roytershteyn%
, Chen%
, Ng%
\BCBL {}\ \BBA {} Hietala%
}{%
Y\BPBI A.~Omelchenko%
\ \protect \BOthers {.}}{%
{\protect \APACyear {2021b}}%
}]{%
omelchenko:2021}
\APACinsertmetastar {%
omelchenko:2021}%
\begin{APACrefauthors}%
Omelchenko, Y\BPBI A.%
, Roytershteyn, V.%
, Chen, L\BHBI J.%
, Ng, J.%
\BCBL {}\ \BBA {} Hietala, H.%
\end{APACrefauthors}%
\unskip\
\newblock
\APACrefYearMonthDay{2021b}{}{}.
\newblock
{\BBOQ}\APACrefatitle {HYPERS simulations of solar wind interactions with the
  Earth's magnetosphere and the Moon} {Hypers simulations of solar wind
  interactions with the earth's magnetosphere and the moon}.{\BBCQ}
\newblock
\APACjournalVolNumPages{Journal of Atmospheric and Solar-Terrestrial
  Physics}{215}{}{105581}.
\PrintBackRefs{\CurrentBib}

\bibitem [\protect \citeauthoryear {%
Omidi%
, Blanco-Cano%
, Russell%
\BCBL {}\ \BBA {} Karimabadi%
}{%
Omidi%
\ \protect \BOthers {.}}{%
{\protect \APACyear {2004}}%
}]{%
omidi:2004}
\APACinsertmetastar {%
omidi:2004}%
\begin{APACrefauthors}%
Omidi, N.%
, Blanco-Cano, X.%
, Russell, C.%
\BCBL {}\ \BBA {} Karimabadi, H.%
\end{APACrefauthors}%
\unskip\
\newblock
\APACrefYearMonthDay{2004}{}{}.
\newblock
{\BBOQ}\APACrefatitle {Dipolar magnetospheres and their characterization as a
  function of magnetic moment} {Dipolar magnetospheres and their
  characterization as a function of magnetic moment}.{\BBCQ}
\newblock
\APACjournalVolNumPages{Advances in Space Research}{33}{11}{1996--2003}.
\PrintBackRefs{\CurrentBib}

\bibitem [\protect \citeauthoryear {%
Omidi%
\ \BBA {} Sibeck%
}{%
Omidi%
\ \BBA {} Sibeck%
}{%
{\protect \APACyear {2007}}%
}]{%
omidi:2007}
\APACinsertmetastar {%
omidi:2007}%
\begin{APACrefauthors}%
Omidi, N.%
\BCBT {}\ \BBA {} Sibeck, D\BPBI G.%
\end{APACrefauthors}%
\unskip\
\newblock
\APACrefYearMonthDay{2007}{}{}.
\newblock
{\BBOQ}\APACrefatitle {Flux transfer events in the cusp} {Flux transfer events
  in the cusp}.{\BBCQ}
\newblock
\APACjournalVolNumPages{Geophysical Research Letters}{34}{4}{}.
\newblock
\begin{APACrefURL}
  \url{https://agupubs.onlinelibrary.wiley.com/doi/abs/10.1029/2006GL028698}
  \end{APACrefURL}
\newblock
\begin{APACrefDOI} \doi{https://doi.org/10.1029/2006GL028698} \end{APACrefDOI}
\PrintBackRefs{\CurrentBib}

\bibitem [\protect \citeauthoryear {%
Palmroth%
\ \protect \BOthers {.}}{%
Palmroth%
\ \protect \BOthers {.}}{%
{\protect \APACyear {2015}}%
}]{%
palmroth:2015}
\APACinsertmetastar {%
palmroth:2015}%
\begin{APACrefauthors}%
Palmroth, M.%
, Archer, M.%
, Vainio, R.%
, Hietala, H.%
, Pfau-Kempf, Y.%
, Hoilijoki, S.%
\BDBL {}Eastwood, J\BPBI P.%
\end{APACrefauthors}%
\unskip\
\newblock
\APACrefYearMonthDay{2015}{}{}.
\newblock
{\BBOQ}\APACrefatitle {ULF foreshock under radial IMF: THEMIS observations and
  global kinetic simulation Vlasiator results compared} {Ulf foreshock under
  radial imf: Themis observations and global kinetic simulation vlasiator
  results compared}.{\BBCQ}
\newblock
\APACjournalVolNumPages{Journal of Geophysical Research: Space
  Physics}{120}{10}{8782-8798}.
\newblock
\begin{APACrefURL}
  \url{https://agupubs.onlinelibrary.wiley.com/doi/abs/10.1002/2015JA021526}
  \end{APACrefURL}
\newblock
\begin{APACrefDOI} \doi{https://doi.org/10.1002/2015JA021526} \end{APACrefDOI}
\PrintBackRefs{\CurrentBib}

\bibitem [\protect \citeauthoryear {%
Palmroth%
\ \protect \BOthers {.}}{%
Palmroth%
\ \protect \BOthers {.}}{%
{\protect \APACyear {2013}}%
}]{%
palmroth:2013}
\APACinsertmetastar {%
palmroth:2013}%
\begin{APACrefauthors}%
Palmroth, M.%
, Honkonen, I.%
, Sandroos, A.%
, Kempf, Y.%
, {von Alfthan}, S.%
\BCBL {}\ \BBA {} Pokhotelov, D.%
\end{APACrefauthors}%
\unskip\
\newblock
\APACrefYearMonthDay{2013}{}{}.
\newblock
{\BBOQ}\APACrefatitle {Preliminary testing of global hybrid-Vlasov simulation:
  Magnetosheath and cusps under northward interplanetary magnetic field}
  {Preliminary testing of global hybrid-vlasov simulation: Magnetosheath and
  cusps under northward interplanetary magnetic field}.{\BBCQ}
\newblock
\APACjournalVolNumPages{Journal of Atmospheric and Solar-Terrestrial
  Physics}{99}{}{41-46}.
\newblock
\begin{APACrefURL}
  \url{https://www.sciencedirect.com/science/article/pii/S1364682612002349}
  \end{APACrefURL}
\newblock
\APACrefnote{Dynamics of the Complex Geospace System}
\newblock
\begin{APACrefDOI} \doi{https://doi.org/10.1016/j.jastp.2012.09.013}
  \end{APACrefDOI}
\PrintBackRefs{\CurrentBib}

\bibitem [\protect \citeauthoryear {%
Phan%
\ \protect \BOthers {.}}{%
Phan%
\ \protect \BOthers {.}}{%
{\protect \APACyear {2003}}%
}]{%
phan:2003}
\APACinsertmetastar {%
phan:2003}%
\begin{APACrefauthors}%
Phan, T.%
, Frey, H\BPBI U.%
, Frey, S.%
, Peticolas, L.%
, Fuselier, S.%
, Carlson, C.%
\BDBL {}Lundin, R.%
\end{APACrefauthors}%
\unskip\
\newblock
\APACrefYearMonthDay{2003}{}{}.
\newblock
{\BBOQ}\APACrefatitle {Simultaneous Cluster and IMAGE observations of cusp
  reconnection and auroral proton spot for northward IMF} {Simultaneous cluster
  and image observations of cusp reconnection and auroral proton spot for
  northward imf}.{\BBCQ}
\newblock
\APACjournalVolNumPages{Geophysical Research Letters}{30}{10}{}.
\newblock
\begin{APACrefURL}
  \url{https://agupubs.onlinelibrary.wiley.com/doi/abs/10.1029/2003GL016885}
  \end{APACrefURL}
\newblock
\begin{APACrefDOI} \doi{https://doi.org/10.1029/2003GL016885} \end{APACrefDOI}
\PrintBackRefs{\CurrentBib}

\bibitem [\protect \citeauthoryear {%
Pi%
, Nemecek%
, Safrankova%
, Grygorov%
\BCBL {}\ \BBA {} Shue%
}{%
Pi%
\ \protect \BOthers {.}}{%
{\protect \APACyear {2018}}%
}]{%
pi:2018}
\APACinsertmetastar {%
pi:2018}%
\begin{APACrefauthors}%
Pi, G.%
, Nemecek, Z.%
, Safrankova, J.%
, Grygorov, K.%
\BCBL {}\ \BBA {} Shue, J\BHBI H.%
\end{APACrefauthors}%
\unskip\
\newblock
\APACrefYearMonthDay{2018}{}{}.
\newblock
{\BBOQ}\APACrefatitle {Formation of the Dayside Magnetopause and Its Boundary
  Layers Under the Radial IMF} {Formation of the dayside magnetopause and its
  boundary layers under the radial imf}.{\BBCQ}
\newblock
\APACjournalVolNumPages{Journal of Geophysical Research: Space
  Physics}{123}{5}{3533-3547}.
\newblock
\begin{APACrefURL}
  \url{https://agupubs.onlinelibrary.wiley.com/doi/abs/10.1029/2018JA025199}
  \end{APACrefURL}
\newblock
\begin{APACrefDOI} \doi{https://doi.org/10.1029/2018JA025199} \end{APACrefDOI}
\PrintBackRefs{\CurrentBib}

\bibitem [\protect \citeauthoryear {%
Pi%
\ \protect \BOthers {.}}{%
Pi%
\ \protect \BOthers {.}}{%
{\protect \APACyear {2017}}%
}]{%
pi:2017}
\APACinsertmetastar {%
pi:2017}%
\begin{APACrefauthors}%
Pi, G.%
, Shue, J\BHBI H.%
, Grygorov, K.%
, Li, H\BHBI M.%
, Nemecek, Z.%
, Safrankova, J.%
\BDBL {}Wang, K.%
\end{APACrefauthors}%
\unskip\
\newblock
\APACrefYearMonthDay{2017}{}{}.
\newblock
{\BBOQ}\APACrefatitle {Evolution of the magnetic field structure outside the
  magnetopause under radial IMF conditions} {Evolution of the magnetic field
  structure outside the magnetopause under radial imf conditions}.{\BBCQ}
\newblock
\APACjournalVolNumPages{Journal of Geophysical Research: Space
  Physics}{122}{4}{4051-4063}.
\newblock
\begin{APACrefURL}
  \url{https://agupubs.onlinelibrary.wiley.com/doi/abs/10.1002/2015JA021809}
  \end{APACrefURL}
\newblock
\begin{APACrefDOI} \doi{https://doi.org/10.1002/2015JA021809} \end{APACrefDOI}
\PrintBackRefs{\CurrentBib}

\bibitem [\protect \citeauthoryear {%
Pitout%
\ \BBA {} Bogdanova%
}{%
Pitout%
\ \BBA {} Bogdanova%
}{%
{\protect \APACyear {2021}}%
}]{%
pitout:2021}
\APACinsertmetastar {%
pitout:2021}%
\begin{APACrefauthors}%
Pitout, F.%
\BCBT {}\ \BBA {} Bogdanova, Y\BPBI V.%
\end{APACrefauthors}%
\unskip\
\newblock
\APACrefYearMonthDay{2021}{}{}.
\newblock
{\BBOQ}\APACrefatitle {The Polar Cusp Seen by Cluster} {The polar cusp seen by
  cluster}.{\BBCQ}
\newblock
\APACjournalVolNumPages{Journal of Geophysical Research: Space
  Physics}{126}{9}{e2021JA029582}.
\newblock
\begin{APACrefURL}
  \url{https://agupubs.onlinelibrary.wiley.com/doi/abs/10.1029/2021JA029582}
  \end{APACrefURL}
\newblock
\APACrefnote{e2021JA029582 2021JA029582}
\newblock
\begin{APACrefDOI} \doi{https://doi.org/10.1029/2021JA029582} \end{APACrefDOI}
\PrintBackRefs{\CurrentBib}

\bibitem [\protect \citeauthoryear {%
Pitout%
, Escoubet%
, Klecker%
\BCBL {}\ \BBA {} R{\`e}me%
}{%
Pitout%
\ \protect \BOthers {.}}{%
{\protect \APACyear {2006}}%
}]{%
pitout:2006}
\APACinsertmetastar {%
pitout:2006}%
\begin{APACrefauthors}%
Pitout, F.%
, Escoubet, C.%
, Klecker, B.%
\BCBL {}\ \BBA {} R{\`e}me, H.%
\end{APACrefauthors}%
\unskip\
\newblock
\APACrefYearMonthDay{2006}{}{}.
\newblock
{\BBOQ}\APACrefatitle {Cluster survey of the mid-altitude cusp: 1. size,
  location, and dynamics} {Cluster survey of the mid-altitude cusp: 1. size,
  location, and dynamics}.{\BBCQ}
\newblock
\BIn{} \APACrefbtitle {Annales Geophysicae} {Annales geophysicae}\ (\BVOL~24,
  \BPGS\ 3011--3026).
\PrintBackRefs{\CurrentBib}

\bibitem [\protect \citeauthoryear {%
Pitout%
, Escoubet%
, Klecker%
\BCBL {}\ \BBA {} Dandouras%
}{%
Pitout%
\ \protect \BOthers {.}}{%
{\protect \APACyear {2009}}%
}]{%
pitout:2009}
\APACinsertmetastar {%
pitout:2009}%
\begin{APACrefauthors}%
Pitout, F.%
, Escoubet, C\BPBI P.%
, Klecker, B.%
\BCBL {}\ \BBA {} Dandouras, I.%
\end{APACrefauthors}%
\unskip\
\newblock
\APACrefYearMonthDay{2009}{}{}.
\newblock
{\BBOQ}\APACrefatitle {Cluster survey of the mid-altitude cusp --; Part 2:
  Large-scale morphology} {Cluster survey of the mid-altitude cusp --; part 2:
  Large-scale morphology}.{\BBCQ}
\newblock
\APACjournalVolNumPages{Annales Geophysicae}{27}{5}{1875--1886}.
\newblock
\begin{APACrefURL} \url{https://angeo.copernicus.org/articles/27/1875/2009/}
  \end{APACrefURL}
\newblock
\begin{APACrefDOI} \doi{10.5194/angeo-27-1875-2009} \end{APACrefDOI}
\PrintBackRefs{\CurrentBib}

\bibitem [\protect \citeauthoryear {%
Pitout%
, Escoubet%
, Taylor%
, Berchem%
\BCBL {}\ \BBA {} Walsh%
}{%
Pitout%
\ \protect \BOthers {.}}{%
{\protect \APACyear {2012}}%
}]{%
pitout:2012}
\APACinsertmetastar {%
pitout:2012}%
\begin{APACrefauthors}%
Pitout, F.%
, Escoubet, C\BPBI P.%
, Taylor, M\BPBI G\BPBI G\BPBI T.%
, Berchem, J.%
\BCBL {}\ \BBA {} Walsh, A\BPBI P.%
\end{APACrefauthors}%
\unskip\
\newblock
\APACrefYearMonthDay{2012}{}{}.
\newblock
{\BBOQ}\APACrefatitle {Overlapping ion structures in the mid-altitude cusp
  under northward IMF: signature of dual lobe reconnection?} {Overlapping ion
  structures in the mid-altitude cusp under northward imf: signature of dual
  lobe reconnection?}{\BBCQ}
\newblock
\APACjournalVolNumPages{Annales Geophysicae}{30}{3}{489--501}.
\newblock
\begin{APACrefURL} \url{https://angeo.copernicus.org/articles/30/489/2012/}
  \end{APACrefURL}
\newblock
\begin{APACrefDOI} \doi{10.5194/angeo-30-489-2012} \end{APACrefDOI}
\PrintBackRefs{\CurrentBib}

\bibitem [\protect \citeauthoryear {%
Plaschke%
\ \protect \BOthers {.}}{%
Plaschke%
\ \protect \BOthers {.}}{%
{\protect \APACyear {2018}}%
}]{%
plaschke:2018}
\APACinsertmetastar {%
plaschke:2018}%
\begin{APACrefauthors}%
Plaschke, F.%
, Hietala, H.%
, Archer, M.%
, Blanco-Cano, X.%
, Kajdi{\v{c}}, P.%
, Karlsson, T.%
\BDBL {}others%
\end{APACrefauthors}%
\unskip\
\newblock
\APACrefYearMonthDay{2018}{}{}.
\newblock
{\BBOQ}\APACrefatitle {Jets downstream of collisionless shocks} {Jets
  downstream of collisionless shocks}.{\BBCQ}
\newblock
\APACjournalVolNumPages{Space Science Reviews}{214}{5}{81}.
\PrintBackRefs{\CurrentBib}

\bibitem [\protect \citeauthoryear {%
Raptis%
, Karlsson%
, Plaschke%
, Kullen%
\BCBL {}\ \BBA {} Lindqvist%
}{%
Raptis%
\ \protect \BOthers {.}}{%
{\protect \APACyear {2020}}%
}]{%
raptis:2020}
\APACinsertmetastar {%
raptis:2020}%
\begin{APACrefauthors}%
Raptis, S.%
, Karlsson, T.%
, Plaschke, F.%
, Kullen, A.%
\BCBL {}\ \BBA {} Lindqvist, P\BHBI A.%
\end{APACrefauthors}%
\unskip\
\newblock
\APACrefYearMonthDay{2020}{}{}.
\newblock
{\BBOQ}\APACrefatitle {Classifying Magnetosheath Jets Using MMS: Statistical
  Properties} {Classifying magnetosheath jets using mms: Statistical
  properties}.{\BBCQ}
\newblock
\APACjournalVolNumPages{Journal of Geophysical Research: Space
  Physics}{125}{11}{e2019JA027754}.
\newblock
\begin{APACrefURL}
  \url{https://agupubs.onlinelibrary.wiley.com/doi/abs/10.1029/2019JA027754}
  \end{APACrefURL}
\newblock
\APACrefnote{e2019JA027754 10.1029/2019JA027754}
\newblock
\begin{APACrefDOI} \doi{https://doi.org/10.1029/2019JA027754} \end{APACrefDOI}
\PrintBackRefs{\CurrentBib}

\bibitem [\protect \citeauthoryear {%
Schwartz%
\ \protect \BOthers {.}}{%
Schwartz%
\ \protect \BOthers {.}}{%
{\protect \APACyear {1992}}%
}]{%
schwartz:1992}
\APACinsertmetastar {%
schwartz:1992}%
\begin{APACrefauthors}%
Schwartz, S\BPBI J.%
, Burgess, D.%
, Wilkinson, W\BPBI P.%
, Kessel, R\BPBI L.%
, Dunlop, M.%
\BCBL {}\ \BBA {} L\"uhr, H.%
\end{APACrefauthors}%
\unskip\
\newblock
\APACrefYearMonthDay{1992}{}{}.
\newblock
{\BBOQ}\APACrefatitle {Observations of short large-amplitude magnetic
  structures at a quasi-parallel shock} {Observations of short large-amplitude
  magnetic structures at a quasi-parallel shock}.{\BBCQ}
\newblock
\APACjournalVolNumPages{Journal of Geophysical Research: Space
  Physics}{97}{A4}{4209-4227}.
\newblock
\begin{APACrefURL}
  \url{https://agupubs.onlinelibrary.wiley.com/doi/abs/10.1029/91JA02581}
  \end{APACrefURL}
\newblock
\begin{APACrefDOI} \doi{10.1029/91JA02581} \end{APACrefDOI}
\PrintBackRefs{\CurrentBib}

\bibitem [\protect \citeauthoryear {%
Shi%
\ \protect \BOthers {.}}{%
Shi%
\ \protect \BOthers {.}}{%
{\protect \APACyear {2020}}%
}]{%
shi:2020}
\APACinsertmetastar {%
shi:2020}%
\begin{APACrefauthors}%
Shi, X.%
, Hartinger, M\BPBI D.%
, Baker, J\BPBI B\BPBI H.%
, Ruohoniemi, J\BPBI M.%
, Lin, D.%
, Xu, Z.%
\BDBL {}Willer, A.%
\end{APACrefauthors}%
\unskip\
\newblock
\APACrefYearMonthDay{2020}{}{}.
\newblock
{\BBOQ}\APACrefatitle {Multipoint Conjugate Observations of Dayside ULF Waves
  During an Extended Period of Radial IMF} {Multipoint conjugate observations
  of dayside ulf waves during an extended period of radial imf}.{\BBCQ}
\newblock
\APACjournalVolNumPages{Journal of Geophysical Research: Space
  Physics}{125}{11}{e2020JA028364}.
\newblock
\begin{APACrefURL}
  \url{https://agupubs.onlinelibrary.wiley.com/doi/abs/10.1029/2020JA028364}
  \end{APACrefURL}
\newblock
\APACrefnote{e2020JA028364 10.1029/2020JA028364}
\newblock
\begin{APACrefDOI} \doi{https://doi.org/10.1029/2020JA028364} \end{APACrefDOI}
\PrintBackRefs{\CurrentBib}

\bibitem [\protect \citeauthoryear {%
Tang%
, Wang%
\BCBL {}\ \BBA {} Li%
}{%
Tang%
\ \protect \BOthers {.}}{%
{\protect \APACyear {2013}}%
}]{%
tang:2013}
\APACinsertmetastar {%
tang:2013}%
\begin{APACrefauthors}%
Tang, B\BPBI B.%
, Wang, C.%
\BCBL {}\ \BBA {} Li, W\BPBI Y.%
\end{APACrefauthors}%
\unskip\
\newblock
\APACrefYearMonthDay{2013}{}{}.
\newblock
{\BBOQ}\APACrefatitle {The magnetosphere under the radial interplanetary
  magnetic field: A numerical study} {The magnetosphere under the radial
  interplanetary magnetic field: A numerical study}.{\BBCQ}
\newblock
\APACjournalVolNumPages{Journal of Geophysical Research: Space
  Physics}{118}{12}{7674-7682}.
\newblock
\begin{APACrefURL}
  \url{https://agupubs.onlinelibrary.wiley.com/doi/abs/10.1002/2013JA019155}
  \end{APACrefURL}
\newblock
\begin{APACrefDOI} \doi{https://doi.org/10.1002/2013JA019155} \end{APACrefDOI}
\PrintBackRefs{\CurrentBib}

\bibitem [\protect \citeauthoryear {%
Taylor%
\ \protect \BOthers {.}}{%
Taylor%
\ \protect \BOthers {.}}{%
{\protect \APACyear {2004}}%
}]{%
taylor:2004}
\APACinsertmetastar {%
taylor:2004}%
\begin{APACrefauthors}%
Taylor, M\BPBI G\BPBI G\BPBI T.%
, Dunlop, M\BPBI W.%
, Lavraud, B.%
, Vontrat-Reberac, A.%
, Owen, C\BPBI J.%
, D\'ecr\'eau, P.%
\BDBL {}Daly, P\BPBI W.%
\end{APACrefauthors}%
\unskip\
\newblock
\APACrefYearMonthDay{2004}{}{}.
\newblock
{\BBOQ}\APACrefatitle {Cluster observations of a complex high-altitude cusp
  passage during highly variable IMF} {Cluster observations of a complex
  high-altitude cusp passage during highly variable imf}.{\BBCQ}
\newblock
\APACjournalVolNumPages{Annales Geophysicae}{22}{10}{3707--3719}.
\newblock
\begin{APACrefURL} \url{https://angeo.copernicus.org/articles/22/3707/2004/}
  \end{APACrefURL}
\newblock
\begin{APACrefDOI} \doi{10.5194/angeo-22-3707-2004} \end{APACrefDOI}
\PrintBackRefs{\CurrentBib}

\bibitem [\protect \citeauthoryear {%
Trattner%
\ \protect \BOthers {.}}{%
Trattner%
\ \protect \BOthers {.}}{%
{\protect \APACyear {2001}}%
}]{%
trattner:2001}
\APACinsertmetastar {%
trattner:2001}%
\begin{APACrefauthors}%
Trattner, K\BPBI J.%
, Fuselier, S\BPBI A.%
, Peterson, W\BPBI K.%
, Chang, S\BHBI W.%
, Friedel, R.%
\BCBL {}\ \BBA {} Aellig, M\BPBI R.%
\end{APACrefauthors}%
\unskip\
\newblock
\APACrefYearMonthDay{2001}{}{}.
\newblock
{\BBOQ}\APACrefatitle {Origins of energetic ions in the cusp} {Origins of
  energetic ions in the cusp}.{\BBCQ}
\newblock
\APACjournalVolNumPages{Journal of Geophysical Research: Space
  Physics}{106}{A4}{5967-5976}.
\newblock
\begin{APACrefURL}
  \url{https://agupubs.onlinelibrary.wiley.com/doi/abs/10.1029/2000JA003005}
  \end{APACrefURL}
\newblock
\begin{APACrefDOI} \doi{https://doi.org/10.1029/2000JA003005} \end{APACrefDOI}
\PrintBackRefs{\CurrentBib}

\bibitem [\protect \citeauthoryear {%
X\BPBI Y.~Wang%
, Lin%
\BCBL {}\ \BBA {} Chang%
}{%
X\BPBI Y.~Wang%
\ \protect \BOthers {.}}{%
{\protect \APACyear {2009}}%
}]{%
wang:2009}
\APACinsertmetastar {%
wang:2009}%
\begin{APACrefauthors}%
Wang, X\BPBI Y.%
, Lin, Y.%
\BCBL {}\ \BBA {} Chang, S\BHBI W.%
\end{APACrefauthors}%
\unskip\
\newblock
\APACrefYearMonthDay{2009}{}{}.
\newblock
{\BBOQ}\APACrefatitle {Hybrid simulation of foreshock waves and ion spectra and
  their linkage to cusp energetic ions} {Hybrid simulation of foreshock waves
  and ion spectra and their linkage to cusp energetic ions}.{\BBCQ}
\newblock
\APACjournalVolNumPages{Journal of Geophysical Research: Space
  Physics}{114}{A6}{}.
\newblock
\begin{APACrefURL}
  \url{https://agupubs.onlinelibrary.wiley.com/doi/abs/10.1029/2008JA013745}
  \end{APACrefURL}
\newblock
\begin{APACrefDOI} \doi{https://doi.org/10.1029/2008JA013745} \end{APACrefDOI}
\PrintBackRefs{\CurrentBib}

\bibitem [\protect \citeauthoryear {%
Y\BPBI L.~Wang%
\ \protect \BOthers {.}}{%
Y\BPBI L.~Wang%
\ \protect \BOthers {.}}{%
{\protect \APACyear {2006}}%
}]{%
wang:2006}
\APACinsertmetastar {%
wang:2006}%
\begin{APACrefauthors}%
Wang, Y\BPBI L.%
, Elphic, R\BPBI C.%
, Lavraud, B.%
, Taylor, M\BPBI G\BPBI G\BPBI T.%
, Birn, J.%
, Russell, C\BPBI T.%
\BDBL {}Zhang, X\BPBI X.%
\end{APACrefauthors}%
\unskip\
\newblock
\APACrefYearMonthDay{2006}{}{}.
\newblock
{\BBOQ}\APACrefatitle {Dependence of flux transfer events on solar wind
  conditions from 3 years of Cluster observations} {Dependence of flux transfer
  events on solar wind conditions from 3 years of cluster observations}.{\BBCQ}
\newblock
\APACjournalVolNumPages{Journal of Geophysical Research: Space
  Physics}{111}{A4}{}.
\newblock
\begin{APACrefURL}
  \url{https://agupubs.onlinelibrary.wiley.com/doi/abs/10.1029/2005JA011342}
  \end{APACrefURL}
\newblock
\begin{APACrefDOI} \doi{https://doi.org/10.1029/2005JA011342} \end{APACrefDOI}
\PrintBackRefs{\CurrentBib}

\bibitem [\protect \citeauthoryear {%
Zhang%
\ \protect \BOthers {.}}{%
Zhang%
\ \protect \BOthers {.}}{%
{\protect \APACyear {2007}}%
}]{%
zhang:2007}
\APACinsertmetastar {%
zhang:2007}%
\begin{APACrefauthors}%
Zhang, H.%
, Dunlop, M\BPBI W.%
, Zong, Q\BHBI G.%
, Fritz, T\BPBI A.%
, Balogh, A.%
\BCBL {}\ \BBA {} Wang, Y.%
\end{APACrefauthors}%
\unskip\
\newblock
\APACrefYearMonthDay{2007}{}{}.
\newblock
{\BBOQ}\APACrefatitle {Geometry of the high-latitude magnetopause as observed
  by Cluster} {Geometry of the high-latitude magnetopause as observed by
  cluster}.{\BBCQ}
\newblock
\APACjournalVolNumPages{Journal of Geophysical Research: Space
  Physics}{112}{A2}{}.
\newblock
\begin{APACrefURL}
  \url{https://agupubs.onlinelibrary.wiley.com/doi/abs/10.1029/2006JA011774}
  \end{APACrefURL}
\newblock
\begin{APACrefDOI} \doi{https://doi.org/10.1029/2006JA011774} \end{APACrefDOI}
\PrintBackRefs{\CurrentBib}

\bibitem [\protect \citeauthoryear {%
Zhang%
\ \protect \BOthers {.}}{%
Zhang%
\ \protect \BOthers {.}}{%
{\protect \APACyear {2013}}%
}]{%
zhang:2013}
\APACinsertmetastar {%
zhang:2013}%
\begin{APACrefauthors}%
Zhang, H.%
, Sibeck, D\BPBI G.%
, Zong, Q\BHBI G.%
, Omidi, N.%
, Turner, D.%
\BCBL {}\ \BBA {} Clausen, L\BPBI B\BPBI N.%
\end{APACrefauthors}%
\unskip\
\newblock
\APACrefYearMonthDay{2013}{}{}.
\newblock
{\BBOQ}\APACrefatitle {Spontaneous hot flow anomalies at quasi-parallel shocks:
  1. Observations} {Spontaneous hot flow anomalies at quasi-parallel shocks: 1.
  observations}.{\BBCQ}
\newblock
\APACjournalVolNumPages{Journal of Geophysical Research: Space
  Physics}{118}{6}{3357-3363}.
\newblock
\begin{APACrefURL}
  \url{https://agupubs.onlinelibrary.wiley.com/doi/abs/10.1002/jgra.50376}
  \end{APACrefURL}
\newblock
\begin{APACrefDOI} \doi{https://doi.org/10.1002/jgra.50376} \end{APACrefDOI}
\PrintBackRefs{\CurrentBib}

\bibitem [\protect \citeauthoryear {%
Zwan%
\ \BBA {} Wolf%
}{%
Zwan%
\ \BBA {} Wolf%
}{%
{\protect \APACyear {1976}}%
}]{%
zwan:1976}
\APACinsertmetastar {%
zwan:1976}%
\begin{APACrefauthors}%
Zwan, B\BPBI J.%
\BCBT {}\ \BBA {} Wolf, R\BPBI A.%
\end{APACrefauthors}%
\unskip\
\newblock
\APACrefYearMonthDay{1976}{}{}.
\newblock
{\BBOQ}\APACrefatitle {Depletion of solar wind plasma near a planetary
  boundary} {Depletion of solar wind plasma near a planetary boundary}.{\BBCQ}
\newblock
\APACjournalVolNumPages{Journal of Geophysical Research
  (1896-1977)}{81}{10}{1636-1648}.
\newblock
\begin{APACrefURL}
  \url{https://agupubs.onlinelibrary.wiley.com/doi/abs/10.1029/JA081i010p01636}
  \end{APACrefURL}
\newblock
\begin{APACrefDOI} \doi{https://doi.org/10.1029/JA081i010p01636}
  \end{APACrefDOI}
\PrintBackRefs{\CurrentBib}

\end{thebibliography}

\end{document}